\documentclass[%
reprint,
 amsmath,amssymb,
 aps,
]{revtex4-1}

\usepackage{graphicx}% Include figure files
\usepackage{dcolumn}% Align table columns on decimal point
\usepackage{bm}% bold math
\usepackage{xcolor,multirow}
\usepackage{hyperref}
\hypersetup{
    colorlinks=false, %set true if you want colored links
    linktoc=all,     %set to all if you want both sections and subsections linked
    citecolor=black,
    filecolor=black,
    linkcolor=black,
    urlcolor=black    
}

\usepackage{float}

\allowdisplaybreaks

\newcommand{\e}{{\rm e}}
\newcommand{\mc}{\mathcal}

\begin{document}

%\preprint{APS/123-QED}

\title{Normative theory of patch foraging decisions}
%another proposed titles (Normative theory of foraging in an uncertain environment)(Bayesian foraging theory in a patchy environment)  

\author{Zachary P Kilpatrick}
 \email{zpkilpat@colorado.edu }
\affiliation{
    Department of Applied Mathematics, University of Colorado, Boulder, Colorado 80309, USA}
 \affiliation{Department of Physiology \& Biophysics, University of Colorado School of Medicine, Aurora, Colorado, USA}

\author{Jacob D Davidson}%
\email{jdavidson@ab.mpg.de}
\affiliation{Department of Collective Behavior, Max Planck Institute of Animal Behavior, Konstanz, Germany}%
\affiliation{Department of Biology, University of Konstanz, Konstanz, Germany}
\affiliation{Centre for the Advanced Study of Collective Behavior, University of Konstanz, Konstanz, Germany}

\author{Ahmed El Hady}
\email{ahady@princeton.edu}
\affiliation{Princeton Neuroscience Institute, Princeton University, Princeton, New Jersey, USA}%

\date{\today}

\begin{abstract}
Foraging is a fundamental behavior as animals' search for food is crucial for their survival. Patch leaving is a canonical foraging behavior, but
%Although there is an increased interest to experimentally study patch foraging decisions,
classic theoretical conceptions of patch leaving decisions lack some key naturalistic details. Optimal foraging theory provides general rules for when an animal should leave a patch, but does not provide mechanistic insights about how those rules change with the structure of the environment. 
Such a mechanistic framework would aid in designing quantitative experiments to unravel behavioral and neural underpinnings of foraging. To address these shortcomings, we develop a normative theory of patch foraging decisions. Using a Bayesian approach, we treat patch leaving behavior as a statistical inference problem. We derive the animals' optimal decision strategies in both non-depleting and depleting environments. A majority of these cases can be analyzed explicitly using methods from stochastic processes.
 %in idealized conditions such as a non-depleting environments and more realistic conditions such as depleting environments including the possibility of returns to previously depleted patches.
 Our behavioral predictions are expressed in terms of the optimal patch residence time and the decision rule by which an animal departs a patch. We also extend our theory to a hierarchical model in which the forager learns the environmental food resource distribution. The quantitative framework we develop will therefore help experimenters move from analyzing trial based behavior to continuous behavior without the loss of quantitative rigor. Our theoretical framework both extends optimal foraging theory and motivates a variety of behavioral and neuroscientific experiments investigating patch foraging behavior.  
%to also include in some way (from discussion:)  The theoretical framework in this work will help experimenters move from trial based behavior to continuous behavior without the loss of quantitative rigor .
\end{abstract}

\pacs{Valid PACS appear here}% PACS, the Physics and Astronomy
\maketitle

\tableofcontents

%Notation: $s$ steps until death, $T$ as total lifetime, $n$ as ring size, $s\leq T \leq ns$.
\section{Introduction}
\label{sec:intro}
Nearly all animals forage, as it is essential to acquire energy for survival through efficient search and harvesting of food resources. 
%Interestingly, the general form of a foraging strategy in spatially-extended environments appears to be fairly universal: animals switch between local search in which they exploit nearby food resources and global search during which they move long distances~\cite{hills2006animal,stephens2008foraging}.
%{need to reword the previous and next sentence:  it sounds like spatial search similar to levy-walks will be a focus.  But instead the point is that many different animals move around and search for food}
%\Acom{I removed the sentence about the spatial aspect of foraging because it is out of context and will confuse readers as we are not treating the spatially extended cases}
Foraging behavior is performed by small organisms like \emph{C. Elegans}~\cite{calhoun2014maximally, Greene_Brown_Dobosiewicz_Ishida_Macosko_Zhang_Butcher_Cline_McGrath_Bargmann_2016} and Drosophila~\cite{sokolowski1984drosophila, landayan2018satiation}, animals with large spatial ranges like birds~\cite{marsh2004energetic, schuck2002rationality}, and mammals like rodents~\cite{arcis2003influence, carter2016rats}, monkeys~\cite{rose1994sex, garber1987foraging}, and humans~\cite{hills2006animal, stephens2008foraging}. 
In addition to its universality across animal species, foraging engages multiple cognitive computations such as learning of food distributions across spatiotemporal scales, statistical inference of food availability,  route planning and decision-making~\cite{hills2006animal}. As such, foraging utilizes multiple neural systems in the brain~\cite{calhoun2015foraging}, offering researchers the opportunity to probe these neural computations simultaneously.   
In addition to studying the proximal neural and behavioral mechanisms shaping foraging across species, one can also ask how these processes and their interplay have been shaped by natural selection to optimize returns in the face of environmental and physiological constraints~\cite{raine2006adaptation, hills2006animal, sokolowski1997evolution, king1986extractive} opening up the opportunity to perform evolutionary quantitative behavioral studies. 
For the aforementioned reasons, there has been an increased interest in studying foraging in a neuroscience context~\cite{mobbs2018foraging, hayden2014neuroscience,hall2019revisiting, calhoun2015foraging}.

Since the space of foraging behaviors is vast, it is crucial to build a conceptual framework starting with a tractable behavior that is sufficiently flexible and rich. Adhering to this aim, patch foraging is both a common and understandable behavior, described as follows~\cite{waage1979foraging,McNamara_1982}: an animal enters a patch of food, harvests resources, and then leaves to search for another patch of food~\cite{stephens2008foraging}. 
%While natural environments are not so cleanly patchy as in this idealization, the assumption that the food resources exist in patches provides a convenient formalism for distinguishing local and global search phases of the forager. 
An animal's behavior can thus be quantified by its patch residence time distribution, travel time distribution, the amount of food resources consumed, and the movement pattern in-between patches. Moreover, the animal's reward rate can be computed by its food intake divided by time. A basic result in behavioral ecology, the marginal value theorem (MVT) states that an animal can optimize energy intake by leaving its current patch when the current reward rate falls below the global average reward rate of the environment~\cite{charnov1976optimal}. This gives a theoretical foundation for assessing optimal decision-making, which has also been validated in many behavioral studies~\cite{waage1979foraging, McNamara_1982, mellgren1982foraging, cuthill1990central, Rita_Ranta_1998, driessen1999patch, ward2000simulation, Eliassen_Jorgensen_Mangel_Giske_2009, taneyhill2010patch, zhang2014recent}.

However, the MVT does not describe mechanistically how an animal uses its foraging experience to learn key environmental features like the distribution of food availability, and assumes the animal already knows the overall average reward rate of the environment. 
Although mechanistic models of foraging have been proposed~\cite{waage1979foraging,McNamara_1982,driessen1999patch,taneyhill2010patch,davidson2019foraging}, these were not derived based on principles of statistical inference, and therefore cannot address how optimal decisions depend on the resource environment and uncertainty in the statistics of food availability.

In behavioral ecology, many studies have considered a Bayesian approach to patch leaving, asking how an animal uses available information to make optimal foraging decisions~\cite{biernaskie2009bumblebees, f2006simpler, mcnamara2006bayes, olsson1998survival,McNamara_Houston_1980,olsson2006bayesian,pierre2008bayesian,olsson1999gaining,van2010state,rodriguez1997density,ellison2004bayesian,mcnamara2006bayes,alonso1995patch,green1980bayesian,olsson1999gaining2,j2006animals,killeen1996bayesian,nonacs1998patch,valone1989measuring}.
However, because most studies consider a specific or narrow range of environmental conditions, it is not clear how a Bayesian approach relates to other mechanistic models of foraging decisions, and how such an approach may assist in deriving possible neural implementations. To date there is not a complete normative theory of how animals make patch leaving decisions under a broad class of environmental conditions.

The aim of our study is thus to develop a Bayesian framework of patch leaving behavior by treating decisions as a statistical inference problem.  Our framework connects normative theory of foraging decisions with mechanistic drift-diffusion models, as proposed in~\cite{davidson2019foraging}.
%We develop a general formal framework that enables the design of behavioral and neural experiments studying foraging behavior taking into account the full richness of behavioral strategies the animal will adopt on different spatio-temporal scales across different environmental statistical structures. By comparing normative patch leaving strategies from a wide variety of environmental contexts side-by-side, we see a number of clear trends and similarities across task conditions.

To realize our theoretical framework, we perform the following model derivations and analysis:
\begin{itemize}
   \item We consider a series of idealizing limits such as non-depleting patches,  homogeneous environments, and binary environments. Using probabilistic sequential updating, we derive stochastic differential equations (SDEs) for the belief about the arrival rate of food in the current patch. This generates analytically tractable models associated with optimal patch leaving strategies.
    \item We consider a series of more naturalistic conditions such as depleting patches, environments in which the forager returns to depleted patches, and environments with many or a continuum of patch types. 
    %In general, we find that the optimal strategies in these more complex environments differ very little from those in idealizing limits.
    \item We derive optimal decision strategies for patch leaving that either maximize the long term reward rate or minimize the time to enter and remain in a high yielding patch across all the aforementioned environmental structures.
    \item We study how animals can learn the distribution of patch types in both non-depleting and depleting environments. This allows us to study how the rate at which the environmental distribution is learned depends on environmental parameters.
\end{itemize}

Following the aforementioned modeling approach, we arrive at a number of general conclusions concerning the best strategies for foragers to adopt in different environments:
\begin{itemize}
    \item In idealizing limits, patch leaving decision statistics are described by solutions of first passage time problems for SDEs with absorbing boundaries.
    \item In non-depleting environments, in which consumption does not diminish food availability, an ideal forager should minimize their time to find and remain in the highest yielding patch in the environment. This time decreases both as high yielding patches become more common and as the highest yielding patch becomes more discriminable.
    \item  In depleting environments, an ideal forager must estimate the current arrival rate of food within a patch and when this becomes too low, depart the patch.
    \item Across a wide range of task parameters, the forager should leave a patch when the current reward rate is matched to the environmental reward rate, as in the MVT. However, if there is a high level of uncertainty about the food availability within a patch, even an ideal forager will tend to stay too long in low yielding patches (overharvesting)  and not long enough in high yielding patches (underharvesting)
    \item When the environment is fully depleting, such that a forager may return again to patches they may have already partially depleted, they can minimize the time to deplete the environment by aiming to fully deplete patches before leaving them.
\end{itemize}

Our aforementioned results highlight that we have found several key deviations from the typical MVT results of classic optimal foraging theory.  These deviations arise due to either limited information available to the forager, slow patch depletion, or full environmental depletion.

Another key aspect of our model that extends beyond classic optimal foraging theory is the ability of our ideal forager to learn the underlying food arrival rates of the environment. The forager learns patch arrival rates more rapidly when rates are high, since food encounters provide more information than times between food encounters. Also, arrival rates are learned more quickly in depleting environments than in non-depleting environments, since successive food encounters rule out low initial arrival rates. Combined with the broad range of optimal foraging strategies derived assuming the observer has full knowledge of its environment, our analysis provides a number of touchstones to be compared with patch leaving statistics from future and past behavioral studies.

%We anticipate that our theoretical framework will facilitate studies of the neural mechanisms associated with naturalistic decision making 
% It is therefore crucial to remember that formal tractability of naturalistic behavioral mechanisms involves facilitating neural mechanistic tractability of those behavior.
We anticipate that our theoretical framework will provide a platform to study behavioral and neural mechanisms of naturalistic decision-making in a similar manner as trained decision-making behavior is studied within systems neuroscience~\cite{mobbs2018foraging}. Our model is therefore useful in opening up the space for laboratory experiments that mimic naturalistic patch foraging dynamics and for generating testable hypothesis under different ecologically relevant experimental designs.

\section{Sequential updating foraging model}
% same as "evidence accumulation".  But use different nomenclature to avoid confusion
\label{sec:model}

\begin{figure*}
\centering \includegraphics[width=17cm]{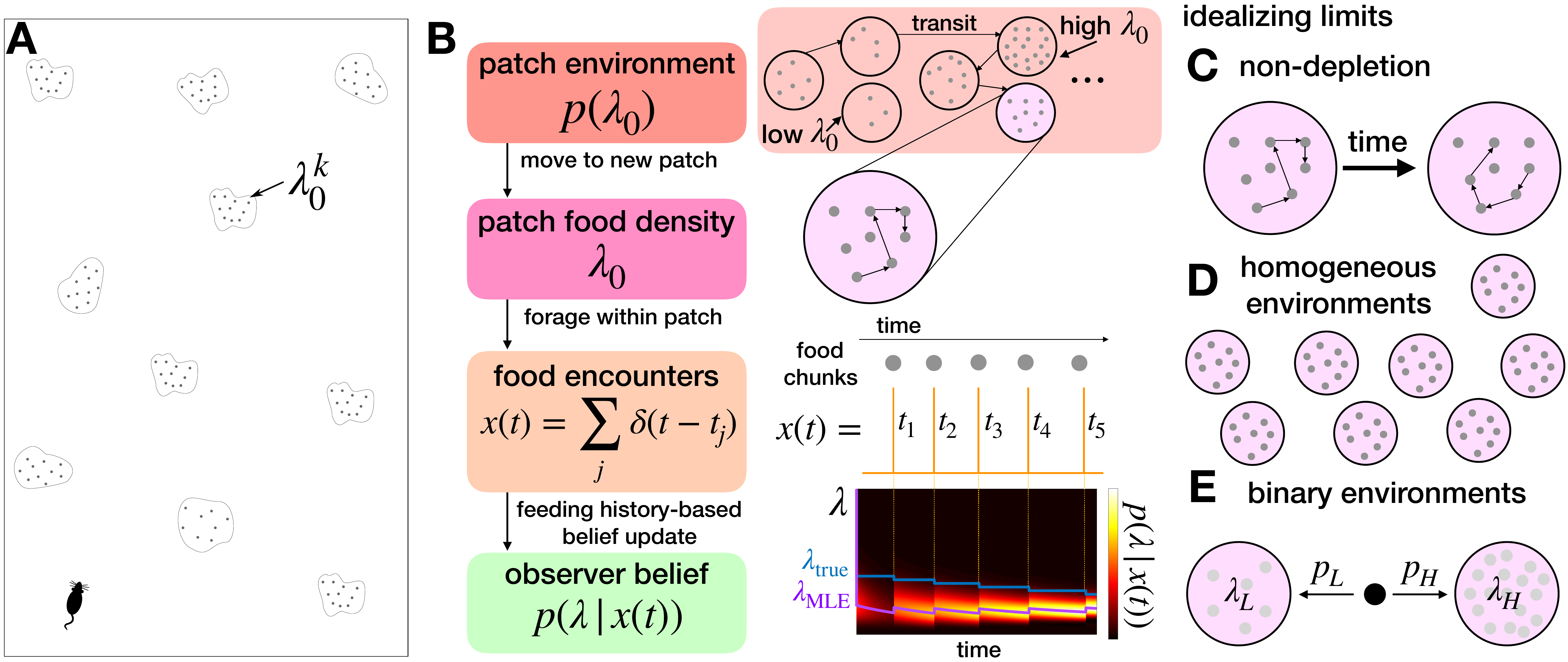}
\caption{\textbf{Patch foraging task and model.} {\bf A.} Task environment: An animal enters an arena and forages from patches, each patch $k$ has some (possibly distinct) initial food yield rate $\lambda_0^k$. {\bf B.} Ideal observer foraging model: Initial patch yield is drawn from the distribution $p(\lambda_0)$, generating random food encounter times $t_{1:K}$, and updating belief of current food arrival rate $\lambda (t)$ for patch. The maximum likelihood estimate $\lambda_{\rm MLE}$ gets approaches the true $\lambda_{\rm true}$ arrival rate over time. {\em Model idealizations:} {\bf C.} Non-depletion: Assuming patches are depleted slowly enough to be approximated as `non-depleting.' {\bf D.} Homogeneous environments: All patches have the same initial yield rate $\lambda_0$. {\bf E.} Binary environment: Two possible initial patch yield rates $\lambda_H > \lambda_L$ occur with probability $p_H$ and $p_L$.}
\label{fig1:schematic}
\end{figure*}

We model an animal searching a large arena representing a natural environment with distributed patches of food (Fig.~\ref{fig1:schematic}A). When the animal enters a food patch, it consumes food within the patch until it decides to leave for another patch. The parameters characterizing the behavior are: environmental distribution of initial food density $p(\lambda_0)$, patch size $A$, food chunk size $c$, and mean travel time between patches $\tau$. In this work, we assume the animal knows the patch size, chunk size, and the travel time (which is held fixed). Note, we could model the learning of these parameters as separate sequential updating processes (See Discussion for more details). Our interest is in how an observer can learn the food density within and across patches over time and use this information to guide an efficient foraging strategy.

\subsection{Full model}
Under simplifying assumptions, we derive a sequential sampling model for an ideal observer's posterior of their current patch's food yield rate $\lambda (t)$. To model the random timing of food encounters, we assume they are Poisson-distributed with a rate $\lambda(t) = \lambda_0 - \rho K(t)$ that decreases with $K(t)$ the number of food chunks found so far, and $\rho$ is the impact of each food chunk on the arrival rate. Thus, the time $t_K$ between the $K$th and $(K+1)$th encounter is drawn from the exponential distribution $(\lambda_0 - \rho K)\e^{- (\lambda_0 - \rho K) t}$. We assume for now that the observer knows and initializes their belief with the prior $p(\lambda_0)$ when arriving in a new patch (Fig.~\ref{fig1:schematic}B). In Section \ref{sec:learn}, we extend our model to consider the process by which the distribution $p(\lambda_0)$ is learned. Given the food encounter sequence $x(t) = K'(t) = \sum_{j=1}^{K(t)} \delta(t-t_j)$, an ideal observer updates their belief about the current arrival rate $\lambda$ according to
\begin{align}
    p(\lambda | x(t)) &= p(x(t)|\lambda) \frac{p_0(\lambda + K(t) \rho)}{p(x(t))} \nonumber \\
    & \propto  \frac{(\lambda/\rho + K)!}{(\lambda/\rho)!} \e^{- \lambda t} p_0(\lambda + K \rho), \ \ \ \lambda \geq 0. \label{fullmod}
\end{align}
Food encounters shape the observer's belief about the current yield rate $\lambda (t)$ in two main ways: (1)~they give evidence of higher yield rates, since encounters are more probable in high yielding patches; and (2)~they deplete the patch, decrementing the yield rate, as described by the $\rho$ terms in the posterior update. Thus, there is a tension between the depleting effects of food encounters, and the evidence they provide for higher yield rates (Fig.~\ref{fig1:schematic}B). 

To better understand how an ideal observer's evidence accumulation process evolves, we consider a few different idealizing limits of Eq.~(\ref{fullmod}). Analyzing these reveals distinct patterns in the belief dynamics as determined by task parameters, as we will show.

\subsection{Idealizing limits}
%\begin{itemize}
%    \item Showing how to get simpler models from most general one.
%\end{itemize}

{\em Non-depletion.} In the limit $\rho \to 0^+$, food encounters do not deplete the rate of food arrival, which could be due to slow search rates or large patch areas~\cite{davidson2019foraging}. Taking this limit, Eq.~(\ref{fullmod}) simplifies considerably to
\begin{align*}
    p(\lambda | x(t)) = \frac{p_0(\lambda)}{p(t_{1:K(t)})} \lambda^{K(t)} \e^{-\lambda t}, \hspace{1cm} \lambda \geq 0, 
\end{align*}
so the log-likelihood update for $\lambda \geq 0$ becomes
\begin{align}
    \log p(\lambda | x(t)) =& K(t) \log \lambda - \lambda t - \log p(t_{1:K(t)}) + \log p_0(\lambda), \label{llnodep}
\end{align}
showing that food encounters provide a pulse of evidence that increases with $\lambda$, and a lack of food results in a linear decrease of the log likelihood of the arrival rate $\lambda$.

In binary environments, with two possible arrival rates $\lambda_H > \lambda_L$, we have $p_0(\lambda) = p_H \delta (\lambda - \lambda_H) + p_L \delta( \lambda - \lambda_L)$ where $p_H + p_L = 1$. Eq.~(\ref{llnodep}) collapses to a scalar update equation for the log-likelihood ratio (LLR),
\begin{align*}
y(t) \equiv& \log \frac{\displaystyle p(\lambda_H|x(t))}{\displaystyle p(\lambda_L | x(t))} \\
=& \underbrace{K(t) \log \frac{\lambda_H}{\lambda_L}}_{\text{food encounters}} - \underbrace{(\lambda_H - \lambda_L)t}_{\text{lack of food}} + \underbrace{\log \frac{p_H}{p_L}}_{\text{prior}},
\end{align*}
which can be written equivalently as a stochastic differential equation (SDE) for the rate-of-change of the LLR:
\begin{align}
    \frac{dy}{dt} = \underbrace{\log \frac{\lambda_H}{\lambda_L}\sum_{j=1}^{\infty} \delta ( t- t_j)}_{\text{food encounters}} - \underbrace{(\lambda_H - \lambda_L)}_{\text{lack of food}}, \label{binnodep}
\end{align}
with initial condition set by the prior $y(0) = \log \frac{\displaystyle p_H}{\displaystyle 1- p_H}$. Eq.~(\ref{binnodep}) has a form similar to classic evidence accumulation models commonly used to model psychophysical data from decision-making tasks~\cite{gold2007neural,brunton2013rats}, as well as to previous models of foraging decisions~\cite{davidson2019foraging}. Due to its simple form, it can be analyzed explicitly to determine how long term statistics are shaped by task parameters.

{\em Homogeneous environments.} Another way of simplifying Eq.~(\ref{fullmod}) is to consider homogeneous environments where knowledge of the patch yield rate is perfect. To see this, note that if $p_0(\lambda) = \delta ( \lambda - \lambda_0)$, then Eq.~(\ref{fullmod}) implies $p(\lambda | x(t)) = \delta (\lambda - (\lambda_0 - K(t) \rho))$. In other words, the observer has perfect knowledge that $\lambda (t) = \lambda_0 - K(t) \rho$.

{\em Binary depleting environments.} We also explore belief updating and patch departure strategies in depleting binary environments, revealing two phases of yield rate inference: an initial discrimination phase followed by depletion based decrementing of the yield rate estimate. In this case, we have $p_0(\lambda) = p_H \delta( \lambda - \lambda_H) + p_L \delta (\lambda - \lambda_L)$, and the associated LLR has a rate-of-change given by the non-autonomous SDE:
%\begin{align*}
%    y(t) \equiv& \log \frac{p(\lambda_H -K(t) \rho| x(t))}{p(\lambda_L - K(t) \rho | x(t))} \\
%    =& \sum_{j=0}^{K(t)-1} \log \frac{\lambda_H - j \rho}{\lambda_L - j \rho} - (\lambda_H - \lambda_L)t + \log \frac{p_H}{p_L},
%\end{align*}
%which has the associated non-autonomous differential equation description
\begin{align}
    \frac{dy}{dt} = \sum_{j=1}^{K_{\rm max}} \log \frac{\lambda_H - (j-1) \rho}{\lambda_L - (j-1) \rho} \cdot  \delta (t - t_j) - (\lambda_H - \lambda_L), \label{bindep}
\end{align}
where $y(0) = \log \frac{\displaystyle p_H}{\displaystyle 1-p_H}$ as in the non-depleting case.

These idealizing limits all afford some level of tractability. However, we first derive optimal patch leaving strategies in non-depleting binary environments. In this case we can explicitly derive first passage time statistics associated with patch departures, allowing us to examine how efficient patch leaving depends on environmental parameters like patch discriminability ($\lambda_H/ \lambda_L$) and high patch prevalence ($p_H$). The trends we find in this idealized case help us develop and interpret optimal strategies in more complex cases like depleting environments and in environments with more than two patch types.

% I put these spaces here due to the odd spacing we were getting at the end of p.4.
\ \\

\ \\

\ \\

\section{Minimizing time to find high patch in non-depleting environments}
\label{nodeplete}

A forager in an environment with non-depleting patches is best off locating a patch with the highest yield and remaining there. In binary environments with two yield rates $\lambda_{H} > \lambda_L$ (Fig.~\ref{fig2:binnodep_scheme}A) and belief update given by Eq.~(\ref{binnodep}), we show an efficient patch departure strategy is to set a single threshold on the LLR (or likelihood) so that when the forager is sufficiently confident they are in a low yielding patch, they leave. We extend this approach to consider environments with multiple patch types (e.g., three or a continuum) and show a similar thresholding strategy can be used for the forager to most quickly find and remain in the highest yielding patches.

\subsection{Binary environment} 

\begin{figure}
    \centering  \includegraphics[width=8.5cm]{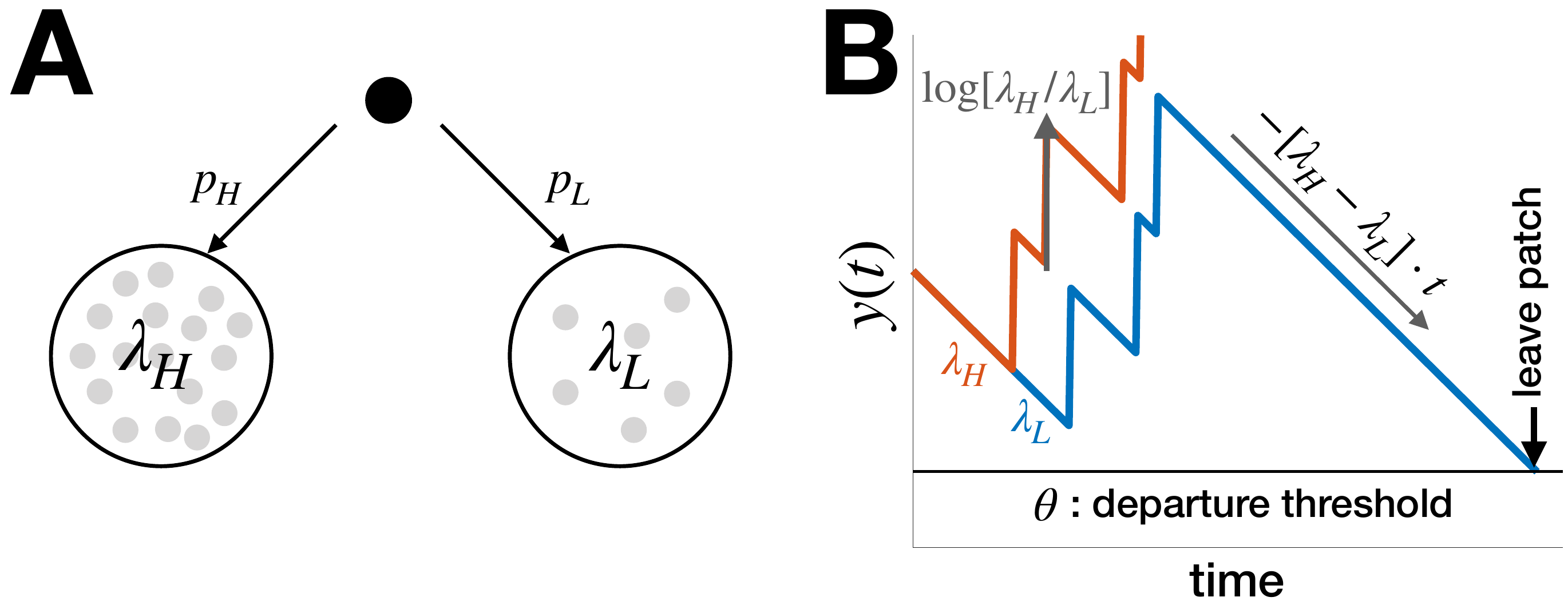} 
    \caption{\textbf{Belief updating in binary non-depleting environments.} {\bf A.} Distribution of high ($\lambda_H$) and low ($\lambda_L$) patch types. {\bf B.} Patch type belief $y(t) = \log \frac{p(\lambda_H|x(t))}{p(\lambda_L|x(t))}$ increases with food encounters and decreases between until $y(t) = \theta$ and the observer departs the patch.}
    \label{fig2:binnodep_scheme}
\end{figure}

Considering Eq.~(\ref{binnodep}), we analyze a formula for the long term reward rate of the forager, and identify optimal parameters for an LLR threshold crossing patch leaving strategy (Fig.~\ref{fig2:binnodep_scheme}B). The reward rate is maximized in the long time limit as long as the forager eventually ends up in a high yielding patch, which occurs via a LLR thresholding strategy. A more thorough analysis of the optimal strategy using dynamic programming could be performed, but as the system does not fit the conditions of decision processes that require time-varying thresholds, we expect this would yield the same result~\cite{drugowitsch2012cost,drugowitsch2014optimal}.

{\em Energy intake rate.} To begin the analysis, define the reward rate function to be maximized over a particular patch leaving policy $\pi$ for a given total foraging time $T$:
\begin{align*}
    R^{\pi}(T) : = \frac{E^{\pi}(T)}{T} = \frac{\langle E_H^{\pi}(T_H , T) \rangle + \langle E_L^{\pi} (T_L, T) \rangle}{T},
\end{align*}
which is the sum of mean food consumed from high and low yielding patch visits $\langle E_{H,L}^{\pi} (T_L, T) \rangle$ divided by the total foraging time $T$. The quantity $E^{\pi}(T)$ is the average number of food chunks consumed. Since policies can condition patch leaving on food chunks consumed, the average food chunk count for a finite time $t$ in a high/low patch will likely not be the mean of the Poisson arrival process $\lambda_{H,L} t$. However, consider the limit $T \to \infty$ of long foraging time:
\begin{align*}
 \lim_{T \to \infty} R^{\pi}(T) = R_H^{\pi}(\infty) + R_L^{\pi}(\infty),   
\end{align*}
in which there are three possible scenarios: (a) the forager spends a nonzero fraction of time in both patches so both $R_{H,L}^{\pi}(\infty) \not\equiv 0$; (b) the forager only spends a nonzero fraction of time in the high yielding patch so $R_L^{\pi}(\infty) \equiv 0$; or (c) the forager only spends a nonzero fraction of time in the low yielding patch so $R_H^{\pi}(\infty) \equiv 0$.

In policies $\pi (a)$ leading to case (a), the forager will spend some fraction of time transitioning between patches and the rest of the time in patches. Define $r_H$ and $r_L$ to be the fraction of time spent in the high and low yielding patches ($r_H + r_L < 1$ due to transits), then
\begin{align*}
    \lim_{T \to \infty} R^{\pi(a)}(T) = r_H \lambda_H + r_L \lambda_L
\end{align*}
for such policies, as the average yield rates $\lambda_{H,L}$ will be recovered in each patch in the long time limit.

On the other hand, any policy which leads to the forager only visiting one patch type for a finite amount of time (a nonzero time fraction as $T \to \infty$) must involve the forager remaining in one patch indefinitely. Otherwise, if they continued patch leaving indefinitely, they would spend a nonzero fraction of time in patches of the other type. As such, policies $\pi(b)$ and $\pi(c)$ causing the forager to (b) stay in the high patch and (c) stay in the low patch imply
\begin{align*}
    \lim_{T \to \infty} R^{\pi(b)}(T) = \lambda_H \ \ \ \text{and} \ \ \ \lim_{T \to \infty} R^{\pi(c)}(T) = \lambda_L
\end{align*}
As the forager's fraction of time spent in the high (low) patch type converges to unity,  $r_{L} \to 0$ ($r_{H} \to 0$) in case b (c). Clearly, all policies of type $\pi(b)$ will maximize long term energy intake rate.

Thus, we can maximize $\lim_{T \to \infty} R^{\pi}(T) $ via any policy whereby the forager finds and remains in a high yielding patch. To do this, we set a lower threshold $\theta < \log  \frac{\displaystyle p_H}{\displaystyle 1-p_H}$ on the belief $y(t)$ in Eq.~(\ref{binnodep}), so the forager leaves given sufficient evidence that the patch is low yielding. We will show subsequently that this leads to a nonzero probability of remaining in a high yielding patch indefinitely each time one arrives there. Thus, there is complete certainty the forager will eventually find and remain in the high patch for all time.

{\em Minimizing time to arrive in high yielding patch.} Given a constant LLR thresholding strategy, we can derive an implicit equation for how the mean time to arrive and remain in the high patch $\bar{T}_{\rm arrive}(\theta)$ depends on $\theta$. This quantity can be computed from the high patch escape probability $\pi_H(\theta)$, the high patch mean visit time $\bar{T}_H(\theta)$ when departing, the low patch mean visit time $\bar{T}_L(\theta)$, the known patch fraction $p_{H}$, and mean transit time $\tau$. Patch departure time statistics can be determined explicitly by solving the mean first passage time problem for the SDE in Eq~(\ref{binnodep}) with absorbing boundary at $y= \theta$. With these quantities in hand, we can compute
\begin{align}
    \bar{T}_{\rm arrive}(\theta) =& \frac{\pi_H(\theta)}{1-\pi_H(\theta)} \left( \bar{T}_H(\theta) + \tau + \frac{1-p_H}{p_H} (\bar{T}_L(\theta) + \tau) \right) \nonumber \\
    & + \frac{1-p_H}{p_H} (\bar{T}_L(\theta) + \tau), \label{tarrive}
\end{align}
where the first term accounts for the number of high patch visits required to remain in the high yielding patch and the second term accounts for the time needed to leave the low yielding patch when starting there. For thresholds further from $y(0) = \log \frac{\displaystyle p_H}{\displaystyle 1-p_H}$, the escape probability $\pi_H(\theta)$ decreases and mean exit times $\bar{T}_{H,L}(\theta)$ increase. Intuitively, if $\theta = \log \frac{\displaystyle p_H}{\displaystyle 1-p_H}$, departure from each patch is immediate, whereas for $\theta \to - \infty$ then $\pi_H(- \infty) \to 0$ but also $\bar{T}_{H,L}(\theta) \to \infty$ since the forager never obtains sufficient evidence to leave a patch. Thus, we expect an interior solution $\theta$ to ${\rm argmin}_{\theta \in (- \infty, \log \frac{p_H}{1-p_H})} \left[ \bar{T}_{\rm arrive}(\theta) \right]$. We now compute the components of Eq.~(\ref{tarrive}) using first passage time methods for SDEs~\cite{gardiner2004handbook}.

{\em Patch departure time statistics.} The problem of finding the time for a forager to leave a patch can be formulated as a first passage time calculation. To obtain explicit results, we derive the corresponding backward Kolmogorov equation of Eq.~(\ref{binnodep}). To do so, first note that the forward Kolmogorov equation is
\begin{align*}
    p_t(y,t) = +ap_y(y,t) + \lambda \Psi (p(y,t),y) - \lambda p(y,t) \equiv {\mc L}p(y,t)
\end{align*}
where $a = \lambda_H - \lambda_L$ and $\lambda \in \{ \lambda_{H}, \lambda_L \}$ is the arrival rate of the current patch, and
\begin{align*}
 \Psi(p(y,t),y) = \left\{ \begin{array}{cc} p(y-b,t), & y>\theta+b, \\ 0, & \theta< y< \theta+b, \end{array} \right.  
\end{align*}
accounts for the fact that jumps in the belief $y$ due to food encounters can only lead to beliefs that are at least one jump length $b = \log \frac{\lambda_H}{\lambda_L}$ from threshold $\theta$. Patch departures are represented by an absorbing boundary condition at the threshold: $p(\theta, t) = 0$. Defining the $L^2$ inner product $\langle u, v \rangle = \int_{\theta}^{\infty} u(y) v(y) dy$ over real functions, we determine the adjoint ${\mc L}^*$ as $\langle {\mc L}u,v \rangle = \langle u, {\mc L}^* v \rangle$, yielding the right hand side of the corresponding backward Kolmogorov equation~\cite{gardiner2004handbook}:
\begin{align}
    q_t &= -aq_y(y',t|y,0) + \lambda q(y',t|y+b,0) - \lambda q(y',t|y,0) \nonumber \\
    & \equiv {\mc L}^*q(y',t|y,0), \label{backkol}
\end{align}
with associated initial and boundary conditions $p(y',0|y,0) = \delta(y'-y)$ for $y,y'>\theta$ and $p(y',t|\theta,0) = 0$ for $t>0$.

Now to obtain patch departure statistics, we can integrate Eq.~(\ref{backkol}) over all possible belief values $y' \in (\theta, \infty)$ to determine the probability that the patch departure time $T$ is after the time $t$. We define the survival function conditioned on the starting belief $y$ as
\begin{align*}
    P(T > t | y) = \int_{\theta}^{\infty} q(y',t|y,0) d y' \equiv G^j(y,t),
\end{align*}
where $j \in \{H,L\}$ indexes the true patch type $\lambda_j$. Upon integrating Eq.~(\ref{backkol}), we find
\begin{align}
    G_t^j = -aG_y(y,t) + \lambda_j \left[ G(y+b,t) - G(y,t) \right] \equiv {\mc L}^*G^j,  \label{Geqn}
\end{align}
on $y \in (\theta, \infty)$, and note that the absorbing boundary condition ensures $G^j(\theta, t) = 0$ for $t>0$. 

\begin{figure*}
    \centering \includegraphics[width=17cm]{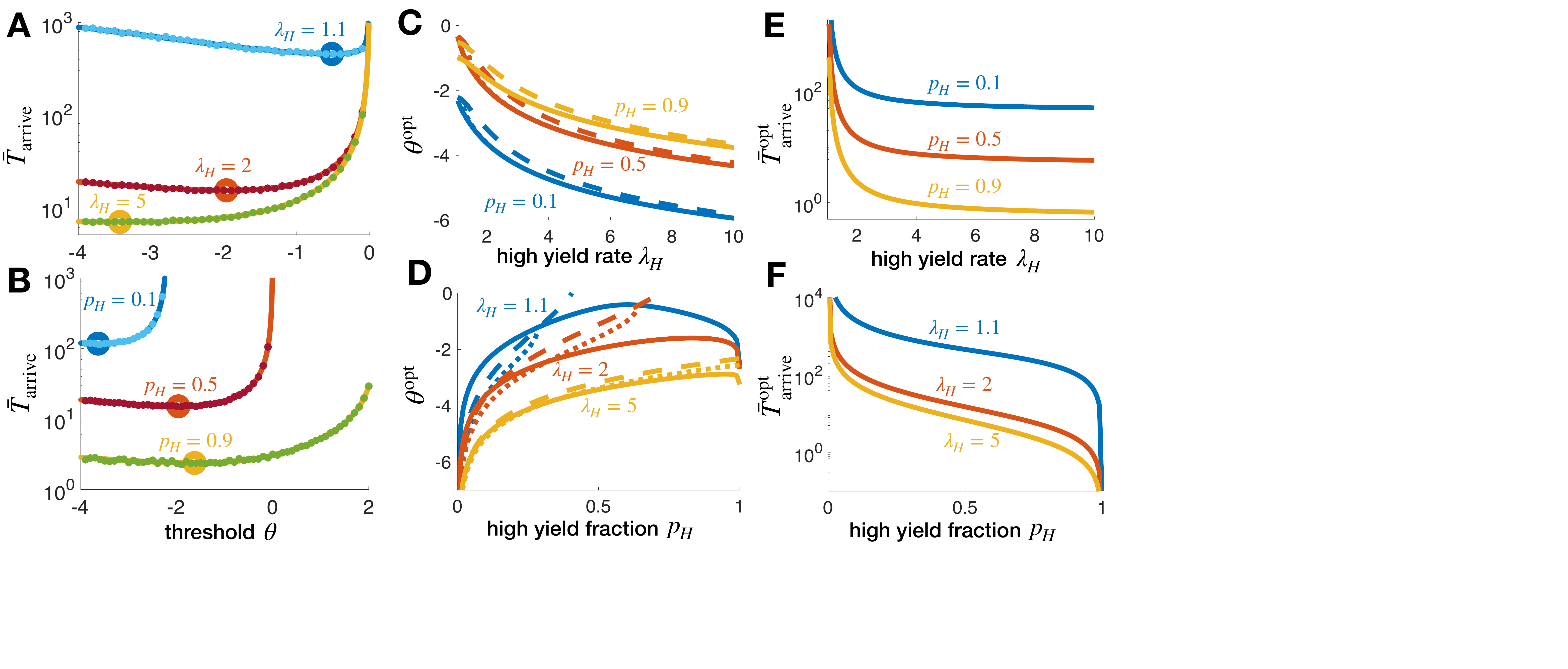} 
    \caption{\textbf{Statistics of high patch identification in binary non-depleting environments.} {\bf A,B.} The mean time to arrive and remain in a high yield patch varies nonmonotonically with departure threshold $\theta$ and decreases as the patch discriminability $\lambda_H/\lambda_L$ and high yield fraction $p_H$ are increased. Solid lines are the analytical result in Eq.~(\ref{tarrive}) and dots are averages over $10^4$ Monte Carlo simulations. $\lambda_H=2$ and $p_H=0.5$ are fixed unless indicated. {\bf C,D.} Departure threshold $\theta^{\rm opt}$ minimizing the time to arrive in the high patch decreases with $\lambda_H/\lambda_L$ and varies nonmonotonically with $p_H$. Solid lines are obtained numerically by solving Eq.~(\ref{fullcritthet}), dotted lines are the Lambert W function approximation (Eq.~(\ref{lamwmin})), and dashed lines are the asymptotic expansion in logarithms (Eq.~(\ref{logexpmin})). {\bf E,F.} The minimal mean time $\bar{T}_{\rm arrive}^{\rm opt}$ to arrive in a high yield patch decreases with  $\lambda_H/\lambda_L$ and $p_H$. We fix $\lambda_L=1$ and $\tau = 5$.}
    \label{fig3:binnodep_stats}
\end{figure*}

Moreover, by taking the limit $t \to 0$ of Eq.~(\ref{Geqn}), we can derive an equation for the escape probability $\pi_j (y) \equiv G^j(x,0) = P(T \geq 0|y)$, which is the probability the belief ever crosses threshold given it begins at $y$ and $t=0$:
\begin{align}
    0 = - a \pi'(y) + \lambda_j \left[ \pi (y+b) - \pi (y) \right] \equiv {\mc L}^*\pi_j(y), \label{pieq}
\end{align}
with absorbing boundary condition $\pi_j (\theta) = 1$. Eigensolutions of Eq.~(\ref{pieq}) take the form $\pi_j (y) = A \e^{ry}$ with corresponding transcendental characteristic equation
\begin{align}
    0 &= -ar + \lambda_j \e^{rb} - \lambda_j = -(\lambda_H - \lambda_L) r + \lambda_j \left( \frac{\lambda_H}{\lambda_L} \right)^r - \lambda_j \nonumber \\
    &= C_j(r). \label{chareqn}
\end{align}
As $\lambda_H > \lambda_L$, $C_j''(r) = \lambda_j b^2 (\lambda_H/ \lambda_L)^{r-2} >0$, this implies $C_j(r)$ is concave up and so has at most two roots: one is $r_1 = 0$, and when $j=H$ ($j= L$) the other root is $r_2 = -1$ ($r_2 = 1$).

To superpose solutions in each case ($j \in \{H,L \}$), note when $j=L$, the average drift
\begin{align*}
    \langle \dot{y} \rangle = \lambda_L \log \frac{\lambda_H}{\lambda_L} - (\lambda_H - \lambda_L) <0
\end{align*}of the belief in Eq.~(\ref{binnodep}) is down toward threshold and we expect $\lim_{y \to \infty} \pi_L (y) = 1$. Applying this and the boundary condition $\pi_L (\theta) = 1$, we see $\pi_L (y) \equiv 1$ when $\lambda = \lambda_L$. However, when $\lambda = \lambda_H$, the average drift
\begin{align*}
    \langle \dot{y} \rangle = \lambda_H \log \frac{\lambda_H}{\lambda_L} - (\lambda_H - \lambda_L) >0
\end{align*}
is away from threshold, implying $\lim_{y \to \infty} \pi_H (y) = 0$ and $\pi_H (\theta) = 1$, yielding $\pi_H(y) = \e^{\theta - y}$.

Now to compute the mean first passage time (MFPT) in either case, note that the probability density function (pdf) $f_j(t)$ of exit times in case $j$ can be calculated from the normalized survival function
\begin{align*}
    \lim_{\Delta t \to 0} \frac{P(T > t|y) - P(T>t+\Delta t|y)}{\Delta t \cdot P(T \geq 0 | y)} = -\frac{G_t^j(y,t)}{\pi_j(y)} \equiv f_j(t).
\end{align*}
The MFPT $\bar{T}_j(y)$ is then obtained by integrating against the pdf:
\begin{align*}
    \bar{T}_j(y) = \int_0^{\infty} t f_j(t) dt = - \frac{\int_0^{\infty} tG_t^j(y,t) dt}{\pi_j (y)} = \frac{\int_0^{\infty} G^j(y,t) dt}{\pi_j (y)}.
\end{align*}
Integrating Eq.~(\ref{Geqn}) over $t \in (0, \infty)$ and defining ${\mc T}_j(y) : = \int_0^{\infty} G^j(y,t) dt = \pi_j (y) \bar{T}_j(y)$, we see
\begin{align}
        - \pi_j(y) = -a {\mc T}_j'(y) + \lambda_j \left[ {\mc T}_j(y+b) - {\mc T}_j(y) \right] \equiv {\mc L}^* {\mc T}_j. \label{bigTeqn}
\end{align}
For $j=L$, the boundary conditions $\bar{T}_L(\theta) = 0$, $\pi_L(\theta) =1$, $\lim_{y \to \infty} \bar{T}_L(y) = \infty$, and $\lim_{y \to \infty} \pi_L(y) = 1$ imply
%where ${\mc T}(\theta) =0$, and we have used the fact that $G(y,0) \equiv \pi (y)$ and $\lim_{t \to \infty} G(y,t) = 0$. Solutions of Eq.~(\ref{bigTeqn}) for $\lambda = \lambda_L$ are found by determining coefficients of ${\mc T}_{L} (y) = A + By$, so
%\begin{align*}
    %-1 = -a B + \lambda_L Bb \ \ \Rightarrow \ \ B = \frac{1}{a - \lambda_L b} = \frac{1}{(\lambda_H - \lambda_L) - \lambda_L \log \frac{\lambda_H}{\lambda_L}},
%\end{align*}
%and the boundary condition ${\mc T}_L (\theta) = 0$ implies $A = - B\theta$, so
\begin{align*}
    \bar{T}_L(y) = \frac{{\mc T}_L(y)}{\pi_L(y)} = \frac{y - \theta}{(\lambda_H - \lambda_L) - \lambda_L \log \frac{\lambda_H}{\lambda_L}}.
\end{align*}
For $j=H$, the boundary conditions $\bar{T}_H(\theta) = 0$, $\pi_H(\theta) =1$, $\lim_{y \to \infty} \bar{T}_H(y) = \infty$, and $\lim_{y \to \infty} \pi_H(y) = 0$ can be leveraged to find an identical expression for the average time in the high patch:
%we use the ansatz ${\mc T}_H(y) = A \e^{-y} + B y \e^{-y}$ and solve for
%\begin{align*}
 %   - \e^{\theta - y} &= -a B \e^{-y} + \lambda_H b B \e^{-y-b} \\
  %  -\e^{\theta} &= - (\lambda_H - \lambda_L) B + \lambda_L \log \frac{\lambda_H}{\lambda_L} B \ \ \Rightarrow \ \ B = \frac{\e^{\theta}}{(\lambda_H - \lambda_L) - \lambda_L \log \frac{\lambda_H}{\lambda_L}},
%\end{align*}
%so applying ${\mc T}_H(\theta) = 0$ implies $A = -B\theta$ and
\begin{align*}
    \bar{T}_H(y) = \frac{{\mc T}_H(y)}{\pi_H(y)} = \frac{y - \theta}{(\lambda_H - \lambda_L) - \lambda_L \log \frac{\lambda_H}{\lambda_L}}.
\end{align*}
For the Bayesian model, the prior and initial condition of a patch is $y(0) = \log \frac{p_H}{1-p_H}$, so the relevant quantities for our problem are $\pi_L \equiv 1$; $\pi_H = \frac{1-p_H}{p_H} \e^{\theta}$; and
\begin{align*}
    \bar{T}_L = \bar{T}_H = \frac{\log \frac{p_H}{1-p_H} - \theta}{(\lambda_H - \lambda_L) - \lambda_L \log \frac{\lambda_H}{\lambda_L}}.
\end{align*}
Substituting these expressions into Eq.~(\ref{tarrive}), we obtain an explicit formula for $\bar{T}_{\rm arrive}$ as it depends on model parameters (Fig.~\ref{fig3:binnodep_stats}A,B). Minimizing $\bar{T}_{\rm arrive}$ to find $\theta^{\rm opt}$ then requires solving:
\begin{align*}
    &\stackrel{\displaystyle {\rm argmin}}{\scriptstyle \theta \in (- \infty, \log \frac{p_H}{1-p_H})}  (1-p_H) \left[ \frac{1+\e^{\theta}}{p_H - (1-p_H) \e^{\theta}} \right] \\ 
    & \hspace{3cm} \times \left[ \frac{\log \frac{p_H}{1-p_H} - \theta}{(\lambda_H - \lambda_L) - \lambda_L \log \frac{\lambda_H}{\lambda_L}} + \tau \right]
\end{align*}
whose solutions $\theta^{\rm opt}$ are identified with critical points obeying $\theta < \log \frac{p_H}{1-p_H}$ and
\begin{align}
  \e^{\theta} \left[ \log \frac{p_H}{1-p_H} - \theta + A \tau \right] = (1+ \e^{\theta}) (p_H - (1-p_H) \e^{\theta}),  \label{fullcritthet}
\end{align}
where $A = (\lambda_H - \lambda_L) - \lambda_L \log \frac{\lambda_H}{\lambda_L}$, which we can solve for $\theta^{\rm opt}$ numerically (solid lines in Fig.~\ref{fig3:binnodep_stats}C,D). An explicit approximation of $\theta^{\rm opt}$ is obtained by dropping the $\e^{2 \theta}$ term (expecting $\theta$ negative) and solving to find
\begin{align}
    \theta^{\rm opt} \approx & W_{-1} \left[ - (1-p_H) \left( \frac{\lambda_H}{\lambda_L} \right)^{\lambda_L \tau} \e^{-(\lambda_H - \lambda_L) \tau} \e^{2p_H-1} \right] \nonumber \\
    & + A \tau + \log \frac{p_H}{1-p_H} +1 -2p_H,  \label{lamwmin}
\end{align}
where $W_{-1}(z)$ is the $(-1)^{\rm th}$ branch of the Lambert $W$ function (inverse of $z = W \e^{W}$). This approximation compares well with numerical solutions to Eq.~(\ref{fullcritthet}) (dotted lines in Fig.~\ref{fig3:binnodep_stats}C,D), and can be further simplified using the approximation $W_{-1}(z) \approx \log(-z) - \log(-\log(-z))$ yielding
\begin{align}
    \theta^{\rm opt} \approx & \log (1-p_H) - \log \left[ -\log(1-p_H)+ (\lambda_H - \lambda_L) \tau \right. \nonumber \\
    & \left. - \lambda_L \tau \log\frac{\lambda_H}{\lambda_L} +1 - 2p_H \right] ,   \label{logexpmin}
\end{align}
from which we can observe how the optimal threshold scales in limits of environmental parameters (dashed lines in Fig.~\ref{fig3:binnodep_stats}C,D). 

Examining Eq.~(\ref{lamwmin}), we see that when high yield patches are rare ($p_H \to 0$), the optimal threshold diverges as $\theta^{\rm opt} \to - \infty$ with the scaling $\log p_H$, since the initial condition becomes more negative. When low yield patches are rare and $p_H \to 1$, $\theta^{\rm opt} \to - \infty$ since the first selected patch will almost always be high yielding. Therefore, if high yielding patches are very rare or common, the threshold should be moved far from zero to require high certainty when abandoning a patch. Between, $\theta^{\rm opt} > - \infty$ varies nonmonotonically in $p_H$. For large $\lambda_H/\lambda_L \gg 1$, $\theta^{\rm opt} \sim - \log (\lambda_H)$, since patches are easier to distinguish as $\lambda_H$ increases, so again high certainty can be required to depart a patch. In the limit $\lambda_H \to \lambda_L$, $\theta^{\rm opt} \approx \log (1-p_H) - \log \left[- \log (1-p_H) + 1-2p_H \right]$. Note, when both increasing discriminability ($\lambda_H/\lambda_L$) and the high yield patch fraction $p_H$, the minimal mean time $T^{\rm opt}_{\rm arrive}$ needed to arrive and remain in a high yield patch decreases (Fig.~\ref{fig3:binnodep_stats}E,F).

Thus, the problem of patch leaving can be reduced to a threshold crossing process in the case of a binary non-depleting environment. Indeed, we find the model is tractable enough to identify explicit scaling relations between model parameters and decision strategies. Namely, the time to arrive and remain in the high yielding patch increases as high patches become less discriminable ($\lambda_H/\lambda_L$ close to one) and more rare (small $p_H$). Patch departures require more certainty when high patches are more discriminable and/or are very rare or common. Next, we extend our study of the high patch arrival problem to non-depleting environments with more than two patch types. Again, we find this can be accomplished by setting a threshold on the belief about whether or not the agent is in the highest yielding patch or not.

\subsection{Three patch types}

\begin{figure*}
    \centering \includegraphics[width=17cm]{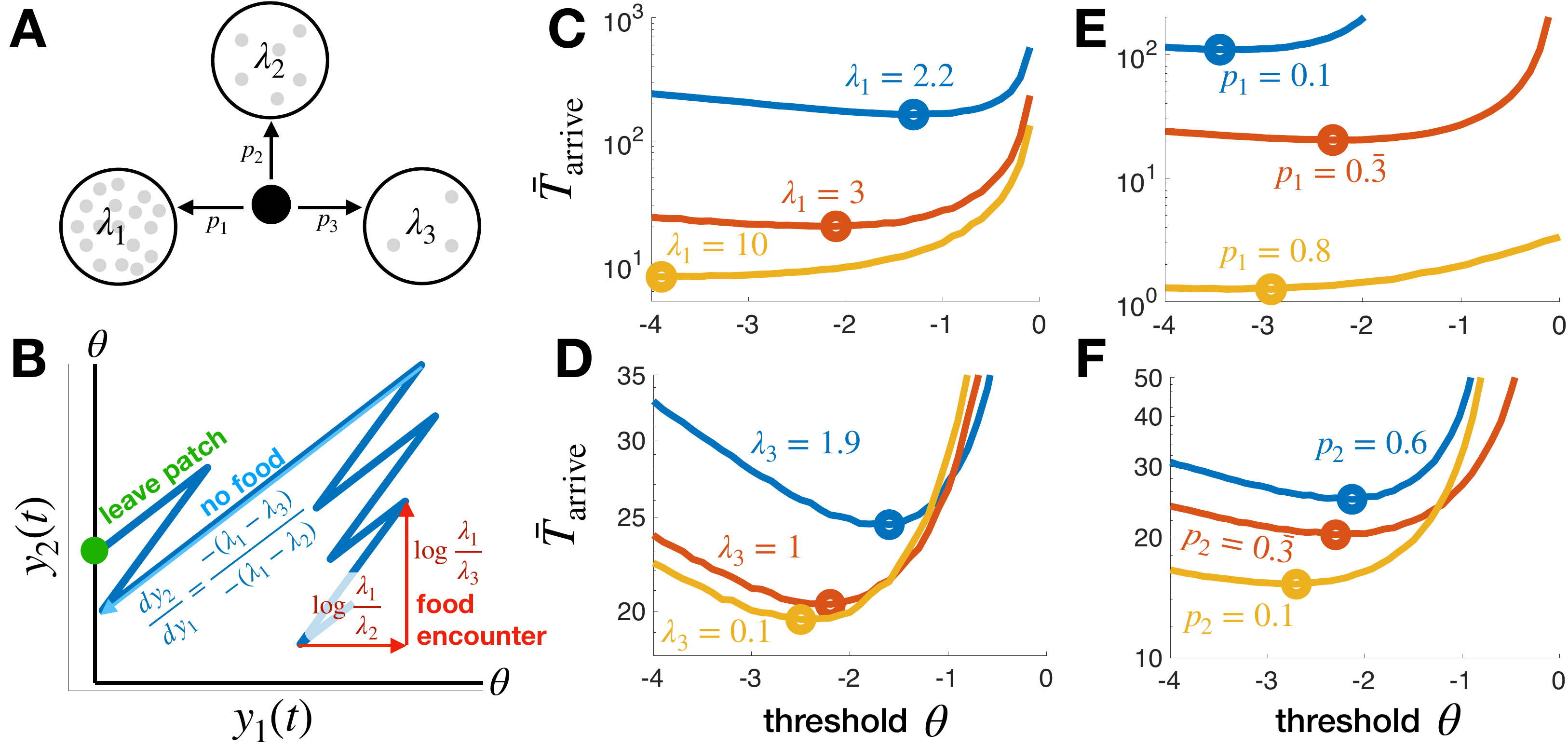}   
    % \includegraphics{}
    % A) distributions of patches.  2-> 'two delta functions'.  3->  more delta fns.  N-> continuous distributions
    % B) Show belief updating process for 3 patches, in 2-dimensional space.  beliefs evolve in 'plane'
    % C) something similar to binary case of fig 2.  don't vary all the parameters, because there will be more.
    \caption{\textbf{Ternary (3 patch) non-depleting environments.} {\bf A.} Distribution of three possible patches. {\bf B.} Patch beliefs $y_1(t) = \log \frac{p(\lambda_1|x(t))}{p(\lambda_2|x(t))}$ and $y_2(t) = \log \frac{p(\lambda_1 | x(t))}{p(\lambda_3|x(t))}$ increase with food encounters and decrease between until either reaches the departure threshold $\theta$. {\bf C.} Mean time $\bar{T}_{\rm arrive}$ to arrive and remain in highest yielding patch $\lambda_1$ (Eq.~(\ref{tarrtern})) decreases with patch discriminability $\lambda_1$ as does the optimal departure threshold $\theta^{\rm opt}$ (circles). Other parameters are $\lambda_2 = 2$, $\lambda_3 = 1$, $\tau = 5$, and $p_1 = p_2=p_3=1/3$. {\bf D.} $\bar{T}_{\rm arrive}$ increases with $\lambda_3$ as does $\theta^{\rm opt}$, since the worst patches become less easy to distinguish from the best $(\lambda_1 = 3)$. Other parameters as before. {\bf E.} $\bar{T}_{\rm arrive}$ decreases with the prevalence of the best patch $p_1$ (while $p_2 = p_3 = (1-p_1)/2$), but $\theta^{\rm opt}$ varies non-monotonically. {\bf F.} $\bar{T}_{\rm arrive}^{\rm opt}$ increases with $p_2$ (while $p_1 = 0.333$ and $p_3 = 1-p_1-p_2$) as does $\theta^{\rm opt}$. All curves for $\bar{T}_{\rm arrive}$ computed from $10^6$ Monte Carlo simulations.}
    \label{fig4:ternenv}
\end{figure*}

In ternary environments, there are three patch arrival rates $\lambda_1 > \lambda_2 > \lambda_3 \geq 0$ with $p_0(\lambda) = \sum_{j=1}^3 p_j \delta (\lambda - \lambda_j)$ and $\sum_{j=1}^3 p_j = 1$ (Fig.~\ref{fig4:ternenv}A). Representing log-likelihoods using Eq.~(\ref{llnodep}) and defining LLRs $y_1 = \log \frac{p(\lambda_1|x(t))}{p(\lambda_2|x(t))}$ and $y_2 = \log \frac{p(\lambda_1 | x(t))}{p(\lambda_3|x(t))}$ for the beliefs within a patch given the food encounter time series $x(t) = \sum_{j=1}^{K(t)} \delta(t-t_j)$, we have the planar system
\begin{subequations} \label{ternnodep}
\begin{align}
    y_1' &= \log \frac{\lambda_1}{\lambda_2} \sum_{j=1}^{K(t)} \delta (t-t_j) - (\lambda_1 - \lambda_2), \\
    y_2' &= \log \frac{\lambda_1}{\lambda_3} \sum_{j=1}^{K(t)} \delta (t-t_j) - (\lambda_1 - \lambda_3),
\end{align}
\end{subequations}
where $y_1(0) = \log \frac{p_1}{p_2}$ and $y_2(0) = \log \frac{p_1}{p_3}$. Note that we can recover any of the three likelihoods from $y_1$ and $y_2$ by the following mappings
\begin{align*}
    p(\lambda_{j+1} | x(t)) &= \frac{\e^{-y_j}}{1 + \e^{-y_1}+\e^{-y_2}},
\end{align*}
%\begin{align*}
 %   p(\lambda_1 | x(t)) &= \frac{1}{1 + \e^{-y_1}+\e^{-y_2}}, \\ p(\lambda_2 | x(t)) &= \frac{\e^{-y_1}}{1 + \e^{-y_1}+\e^{-y_2}}, \\ p(\lambda_3 | x(t)) &= \frac{\e^{-y_2}}{1 + \e^{-y_1}+\e^{-y_2}}.
%\end{align*}
where $y_0=0$ for $j=0$. An argument similar to the binary case demonstrates that the best patch leaving policies in ternary environments cause the forager to find and remain in the highest yielding patch ($\lambda_1$). This is accomplished by thresholding the probability of being in the high yielding patch, so when $p(\lambda_1 | x(t)) = \phi \in (0,p_1)$, the forager exits the patch. In $(y_1,y_2)$ space, this threshold is a parameterized curve
\begin{align}
    y_2^{\phi} = - \log \left[ \frac{1-\phi}{\phi} - \e^{-y_1^{\phi}} \right], \ \ \ y_1^{\phi} \in ( - \log \frac{1-\phi}{\phi}, \infty), \label{terncurvebound}
\end{align}
bounded by $\lim_{y_j^{\phi} \to \infty} y_k^{\phi} = - \log \frac{1-\phi}{\phi} \equiv \theta$, suggesting we approximate the boundary in Eq.~(\ref{terncurvebound}) with $y_1, y_2 \geq \theta$. This leads to the forager departing the patch given sufficient evidence they are not in the highest yielding patch (Fig.~\ref{fig4:ternenv}B).

\begin{figure*}
    \centering
    \includegraphics[width=17cm]{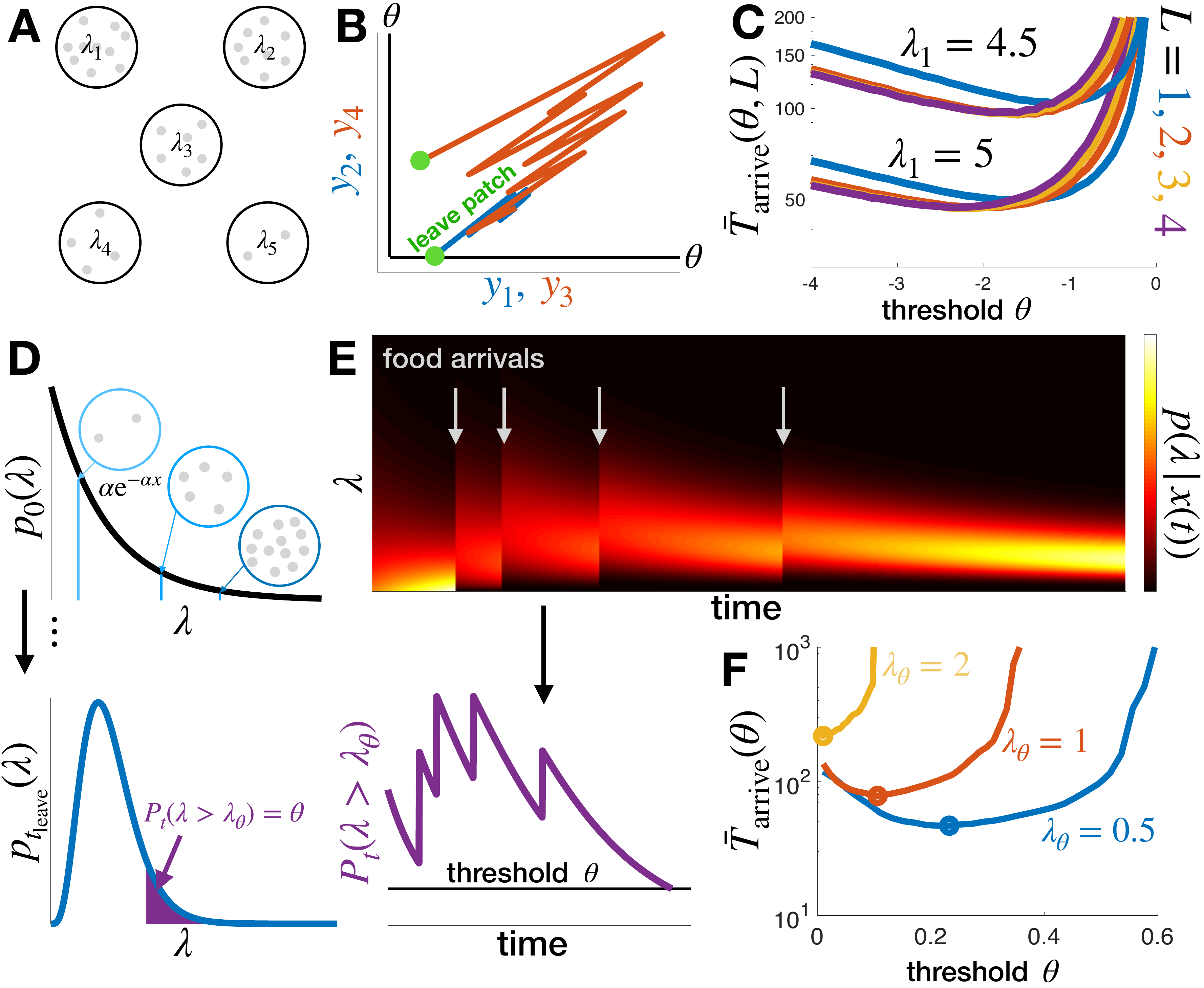}
    % A) distributions of patches.  N-> continuous distributions
    % B) Some results of N patches
    \caption{\textbf{Finding high yielding patches among many patch types.} {\bf A.}~Environment with 5 patch types. {\bf B.}~Patch beliefs $y_j(t) = \log \frac{p(\lambda_1 | x(t))}{p(\lambda_{j+1}|x(t))}$ evolve on an $N-1$-dimensional space until one $y_j$ ($j=1,...,L \leq N-1$) reaches $\theta$. {\bf C.}~The mean time $\bar{T}_{\rm arrive}$ to arrive and remain in highest yielding patch $\lambda_1$ as computed in Eq.~(\ref{tarrNenv}) slightly decreases as more LLRs ($L$ increases) are used ($p_j = 1/N$, $\lambda_j = 6-j$, $j=1,2,3,4,N=5$). {\bf D.}~Continuum of patch types given by an exponential prior $\lambda \sim {\exp}[\alpha]$. The forager departs a patch when the probability that the arrival rate is above threshold ($\lambda > \lambda_{\theta}$) falls to a threshold value $P_t(\lambda > \lambda_{\theta}) = \theta$. {\bf E.}~The posterior $p(\lambda | x(t))$ of the patch arrival rate is shifted up by food arrives, while it decreases in the time between. The mass $P_t(\lambda > \lambda_{\theta})$ follows the same pattern. {\bf F.}~There is an optimal $\theta$ that minimizes the time to arrive and remain in a high yielding patch ($\lambda > \lambda_{\theta}$), and the optimal time $\bar{T}_{\rm arrive}^{\rm opt}$ (circles) increases as the acceptable arrival rate is increased. $\alpha =1$ is used here. In {\bf C} and {\bf F}, $10^7$ Monte Carlo simulations are used to compute the curves $\bar{T}_{\rm arrive}$.}
    \label{fig5:moreno}
\end{figure*}

Given such a strategy, the mean time $\bar{T}_{\rm arrive}$ to arrive and remain in the high yielding patch is computed from the escape probability $\pi_1(\theta)$ from the high yielding patch and the mean times to visit each patch $\bar{T}_j(\theta)$ ($j=1,2,3$) when escaping. As before, the forager will always escape the lower yielding patches (2 and 3). The mean arrival time is then given by
\begin{align}
    \bar{T}_{\rm arrive}(\theta) =& \frac{\pi_1(\theta)}{1- \pi_1(\theta)} \left( \bar{T}_1(\theta)  + \tau \right) \label{tarrtern} \\
    & + \frac{1-p_1}{p_1} \frac{p_2(\bar{T}_2(\theta) + \tau)+p_3(\bar{T}_3(\theta)+\tau)}{1 - \pi_1(\theta)}.  \nonumber
\end{align}
%To optimize this function, we compute the statistics of exit times from each patch, given the policy we have defined below.
%\noindent
%{\em Patch departure statistics.} As in binary environments, we can leverage backward Kolmogorov equations of Eq.~(\ref{ternnodep}) to formulate the mean first passage time problem in ternary environments (See Appendix~\ref{app:ternno}). 
% put 'reason' for using sampling in appendix.  here it sounds like a shortcoming of the work, but its not
While we could compute the needed MFPT statistics for Eq.~(\ref{tarrtern}) by solving a corresponding backward Kolmogorov equation, the resulting two-dimensional delayed PDE proves more computationally expensive to solve than Monte Carlo sampling Eq.~(\ref{ternnodep}). We thus save a more detailed asymptotic analysis of higher order PDEs associated with multi-patch problems for future work, and report results from sampling here (Fig.~\ref{fig4:ternenv}C-F). The mean high patch arrival time $\bar{T}_{\rm arrive}$ depends strongly on the high patch food arrival rate $\lambda_1$, decreasing considerably as the patch becomes more discriminable (Fig.~\ref{fig4:ternenv}C). On the other hand, $\bar{T}_{\rm arrive}$ depends weakly on the worst patch's food arrival rate $\lambda_3$ (Fig.~\ref{fig4:ternenv}D).  % note the linear scale in Fig 4D, compared to log scale in 4C
The discriminability of $\lambda_2$ and $\lambda_3$ does not strongly affect the observer's ability to determine whether a patch is high yielding ($\lambda_1$) or one of the two lower yielding patches ($\lambda_2$ or $\lambda_3$) as long as $\lambda_1$ and $\lambda_2$ are sufficiently discriminable.
%This is because departure depends chiefly on how the beliefs $y_1$ and $y_2$ differentiate the highest patch from the other two.
In a related way, $\bar{T}_{\rm arrive}$ is much more strongly affected by changes in the fraction of the high yielding patch ($p_1$: Fig.~\ref{fig4:ternenv}E) than changes in the balance of the middle ($\lambda_2$) and low ($\lambda_3$) patches (Fig.~\ref{fig4:ternenv}F).

Importantly, the optimal threshold $\theta^{\rm opt}$ decreases as patches become more easily discriminable (Fig.~\ref{fig4:ternenv}C,D), but can vary non-monotonically with the fraction $p_1$ of high patches as in the binary environment (Fig.~\ref{fig3:binnodep_stats}D). Increases in $\theta^{\rm opt}$ at lower values of $p_1$ arise from the fact that departure from high valued patches will be interspersed with long sequences of visits to lower yielding patches. Thus, there is a higher premium on distinguishing the high yielding patch when actually in one, similar to how subjects raise their decision thresholds when intertrial intervals are lengthened in two alternative forced choice paradigms~\cite{bogacz2010humans}. At higher values of $p_1$, one is more likely to land in a high yielding patch, so again one can afford to require more certainty to depart.

We now extend our analysis of departure strategies to environments with $N>3$ patch types. As before, we expect that patch discrimination will mainly be impacted by the spacing between the highest and second highest food arrival rates $\lambda_1$ and $\lambda_2$, as we now show in detail.

\subsection{More patch types}

Consider environments with $N$ patch types having arrival rates $\lambda_1 > \lambda_2> \cdots > \lambda_N \geq 0$ with $p_0(\lambda) = \sum_{j=1}^N p_j \delta (\lambda - \lambda_j)$ and $\sum_{j=1}^{N} p_j = 1$ (Fig.~\ref{fig5:moreno}A). In this case, defining the LLRs $y_j = \log \frac{p(\lambda_1 | x(t))}{p(\lambda_{j+1}|x(t))}$ for $j=1,...,N-1$, yields the $N-1$-dimensional system
\begin{align}
   y_j' = \log \frac{\lambda_1}{\lambda_{j+1}} \sum_{j=1}^{\infty} \delta (t- t_j) - (\lambda_1 - \lambda_{j+1}),  \label{Nnodepllr}
\end{align}
where $y_j(0) = \log \frac{p_1}{p_{j+1}}$, and any likelihood can be recovered as
\begin{align*}
    p(\lambda_{j+1} | x(t)) = \frac{\e^{-y_j}}{1 + \sum_{k=1}^{N-1} \e^{-y_k}},
\end{align*}
where $y_0 = 0$ for $j=0$.

A full analysis of optimal strategies across a broad range of parameters is prohibitive, so we focus on the impact of varying a single departure threshold $\theta$ on the time to arrive and  remain in a high yielding patch. Noting the weak dependence of arrival times on parameters for the low patch ($p_3$ and $\lambda_3$) in the three patch case, we consider simplified strategies in which the observer only tracks the first $L$ LLRs $y_1, y_2, ..., y_L$ and compares these with the threshold $\theta$ to decide when to leave the patch (Fig.~\ref{fig5:moreno}B). Thus, we compute the mean time to arrive and remain in the high patch:
\begin{align}
    \bar{T}_{\rm arrive}(\theta) =& \frac{\pi_1(\theta, L)}{1 - \pi_1(\theta, L)} \left( \bar{T}_1(\theta, L) + \tau \right) \label{tarrNenv} \\
    & + \frac{1-p_1}{p_1} \frac{\sum_{j=2}^N p_j(\bar{T}_j(\theta, L) + \tau)}{1- \pi_1(\theta)}, \nonumber
\end{align}
where patch departure strategy depends on the number $L$ of LLRs thresholded and the threshold $\theta$ used. In a five patch type example, performance depends weakly on $L>2$, and it is sufficient to simply track the LLRs between the first three patches (Fig.~\ref{fig5:moreno}C).

\subsection{Continuum limit}

Next, consider the limit of many patches, in which patch arrival rate $\lambda$ is drawn from a continuous distribution $p_0(\lambda)$ defined on $\lambda \in [0, \infty)$ (Fig.~\ref{fig5:moreno}D).  This general case is described in Eq.~(\ref{llnodep}). Rather than resorting to LLRs, we retain the posterior $p(\lambda | x(t))$, and compute the mass of a thresholded portion of this pdf. Given a reference arrival rate $\lambda_{\theta}$, we track $P(\lambda > \lambda_{\theta} | x(t)) = \int_{\lambda_{\theta}}^{\infty} p(\lambda|x(t)) d \lambda$. This is analogous to a forager who seeks patches with yield rates $\lambda_{\theta}$ or above, but deems lower yield rates to be insufficient. For any continuous pdf $p_0(\lambda)$, the maximum $\lambda$ will never be sampled, so arriving and remaining in the true maximum yielding patch is infeasible.

For the case of an exponential prior $p_0(\lambda) = \alpha \e^{- \alpha \lambda}$, given $K(t)$ food arrivals, we can compute
\begin{align*}
    P(\lambda > \lambda_{\theta} | x(t)) & = \frac{1}{p(x(t))} \int_{\lambda_{\theta}}^{\infty} \alpha \e^{- \alpha \lambda} \lambda^{K(t)} \e^{- \lambda t} d \lambda \\
    &= \frac{\alpha}{(\alpha + t)^{K(t)+1}p(x(t))} \int_{(\alpha + t) \lambda_{\theta}}^{\infty} \nu^{K(t)} \e^{- \nu} d \nu \\
    &= \frac{\alpha}{(\alpha + t)^{K(t)+1}p(x(t))} \Gamma (K(t),(\alpha + t) \lambda_{\theta}),
\end{align*}
where $\Gamma(s,x)$ is the incomplete gamma function, and in a similar way
\begin{align*}
    P(\lambda < \lambda_{\theta} | x(t)) = \frac{\alpha}{(\alpha + t)^{K(t)+1} p(t_{1:K(t)})} \gamma (K(t),(\alpha + t) \lambda_{\theta}),
\end{align*}
where $\gamma (s,x)$ is the lower incomplete gamma function. As such, we can compute the LLR
\begin{align}
    \rho(t) = \log \frac{P(\lambda > \lambda_{\theta}|x(t))}{P(\lambda < \lambda_{\theta} | x(t))} = \log \frac{\Gamma (K(t), (\alpha +t) \lambda_{\theta})}{\gamma (K(t), (\alpha + t) \lambda_{\theta})}, \label{rhocont}
\end{align}
and state the forager departs the patch when $ \rho (t) \leq \hat{\theta} $ or when $P(\lambda > \lambda_{\theta}|x(t)) \leq \theta : = 1/(1 + \e^{- \hat{\theta}})$ (Fig.~\ref{fig5:moreno}D,E). Note, to allow evidence accumulation, we require that $\theta < \alpha \int_{\lambda_{\theta}}^{\infty} \e^{-\alpha \lambda} d \lambda = \e^{- \alpha \lambda_{\theta}} \equiv \phi$, which represents the fraction of patches where $\lambda \geq \lambda_{\theta}$.

%For instance, assume the forager wishes to arrive and remain in one of the top $\phi$-fraction quality patches. From the exponential prior $p_0(\lambda) = \alpha \e^{- \alpha \lambda}$, these are patches with arrival rate $\lambda \geq \lambda_{\rm min}$, where $\alpha \int_{\lambda_{\rm min}}^{\infty} \e^{- \alpha \lambda} d \lambda = \e^{- \alpha \lambda_{\rm min}} = \phi$ or $\lambda_{\rm min} = - (\log \phi)/\alpha $. Thus, for the patch leaving rule, we must set $\lambda_{\theta} \leq \lambda_{\rm min}$ and $\theta < \e^{- \alpha \lambda_{\theta}}$ as shown above.

As with other cases, we can compute the mean time to find and remain in a patch of high enough quality. In fact, we can marginalize over all patch types of each class (high and low) to compute the probability of escaping a high patch across all high patches,
\begin{align*}
    \pi_H(\theta; \lambda_{\theta}) = \int_{\lambda_{\theta}}^{\infty} p_0(\lambda) \pi(\theta; \lambda) d \lambda ,
\end{align*}
and the mean time per visit to a high and low patch types,
\begin{align*}
    \bar{T}_H(\theta; \lambda_{\theta}) &= \int_{\lambda_{\theta}}^{\infty} p_0(\lambda) \bar{T} (\theta; \lambda) d \lambda, \\
    \bar{T}_L(\theta; \lambda_{\theta}) &= \int_{0}^{\lambda_{\theta}} p_0(\lambda) \bar{T}(\theta; \lambda) d \lambda,
\end{align*}
when departing across all patches of each type. We compute the integrals above via Monte Carlo sampling. Defining $p_H = \int_{\lambda_{\theta}}^{\infty} p_0(\lambda) d \lambda$, the mean arrival time is then given by the two patch formula, Eq.~(\ref{tarrive}), as we have partitioned the environment into two patch categories.

As the observer places a higher threshold $\lambda_{\theta}$ on the quality of an acceptably high yielding patch, the optimal time to arrive in such a patch increases (Fig.~\ref{fig5:moreno}F). Moreover, the optimal threshold $\theta$ decreases, as more time must be spent in patches to properly discriminate a high yielding patch, as these become rarer in the prior as $\lambda_{\theta}$ increases. In connection with the binary environment, increasing $\lambda_{\theta}$ corresponds to making high patches more discriminable but also more rare.
%to formulate an update equation for the LLRs:
%\begin{align}
%    \frac{d y(\lambda, t)}{dt} = \log \frac{\lambda}{\lambda_r} \sum_{j=1}^{\infty} \delta (t - t_j) - (\lambda - \lambda_r),
%\end{align}
%where $\lambda \in [0, \infty)$ and $\lambda_r \in [0, \infty)$ is a reference value of $\lambda$ of interest to a given patch leaving strategy. One possibility is to take $\lambda_r : = \bar{\lambda} = \int_0^{\infty} \lambda p_0(\lambda) d \lambda$, so the reference value is the mean of the prior. In this case, all LLRs $y(\lambda,t)$ with $\lambda < \bar{\lambda}$ increase between food encounters and all LLRs with $\lambda > \bar{\lambda}$ increase at food encounters, and vice versa. We can always apply a change of variables to obtain:
%\begin{align*}
%    p(\lambda|x(t)) = \frac{\e^{y(\lambda,t)}}{\int_0^{\infty}\e^{y(\lambda',t)} d \lambda' }.
%\end{align*}
%This allows us to invoke a policy so that when $P(\lambda>\lambda_{\theta}|x(t)) < \theta$, the forager departs the patch, for some minimal acceptable arrival rate $\lambda_{\theta}$. 

Our analysis of patch departure strategies in the context of non-depleting patches reveals a number of consistent trends across patch type counts. First, the optimal time to arrive and remain in the highest yielding patch decreases as the high patch discrimability increases and as high patches become more common. Second, in environments with more than two patch types, the most important parameters in determining the time to find the highest yielding patch are those related to the highest and second highest yielding patch types. To most efficiently find the high patch, it is sufficient to compute LLRs corresponding to the first two or three patch types. In this regard, we expect that reasonably effective strategies for foraging environments with a continuum of patch types could be generated using particle filters that only compute likelihoods over a finite sample of possible patch types~\cite{glaze2018bias}. Strategies for non-depleting environments involve minimizing the time to find a high yielding patch. In depleting environments, the forager eventually must leave any patch it arrives in, as food sources are exhausted. We study such strategies in the next section.

%\begin{itemize}
%\item Another figure perhaps showing how strategies change with increasing number, spread of patch types, in parallel with discriminability and high yield bias of different environments.
%\item Hopefully, we find some general principle of how to set patch leaving threshold based on features of environmental distribution (and then resultant patch leaving time distributions).
%\end{itemize}

\section{Depletion- vs. uncertainty-driven decisions in depleting environments}
\label{deplete}

Foraging animals deplete resources as they use them~\cite{cuthill_starlings_1990,outreman_effects_2005}. This is accounted for in the most general form of the foraging agent's belief in Eq.~(\ref{fullmod}), and thus far we have analyzed idealized versions of the model that neglect this. To understand the impact of depletion on the agent's belief and strategy, we will focus here on simple cases: (a) homogeneous environments in which all patches are the same and the forager knows the initial arrival rate $\lambda_0$ a priori; (b) binary environments in which the forager knows there are two types of patches, and may or may not have to infer which of the two they are currently foraging in while they deplete it; and (c) environments with a few patches in which there is a memory of depletion from previous patch visits.

\subsection{Homogeneous environments}

\begin{figure*}
    \centering
    \includegraphics[width=17cm]{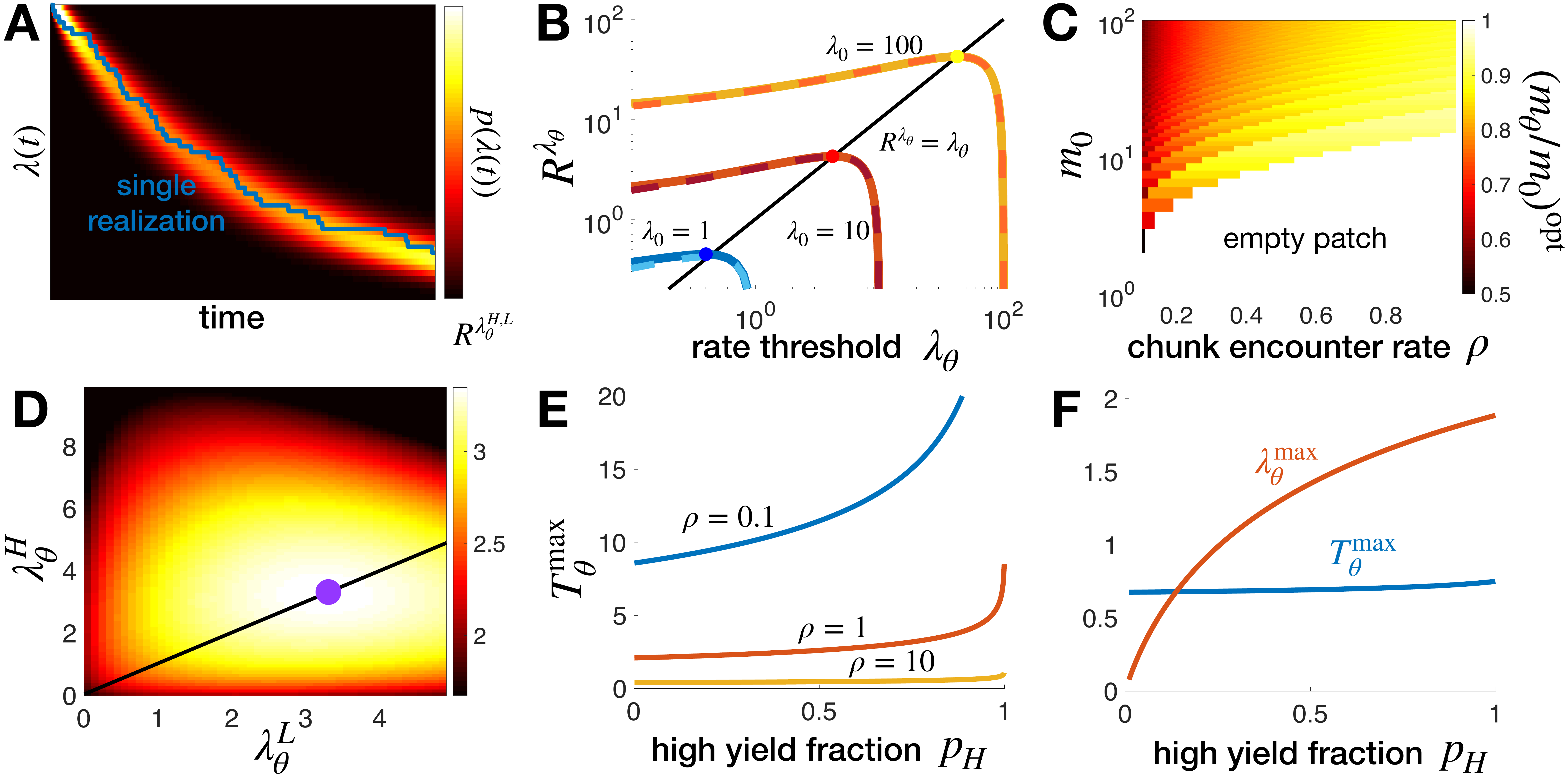}
    \vspace{-4mm}
    \caption{\textbf{Patch leaving strategies in a depleting environment.} {\bf A.} In homogeneous environments where all patches have initial arrival rate $\lambda_0$, the forager maintains full knowledge of the current arrival rate by counting food encounters. {\bf B.} Rate of food consumption as a function of strategy $R^{\lambda_{\theta}}$, where the forager departs once arrival rate falls to or below $\lambda_{\theta}$. Solid lines are exact solution of Eq.~(\ref{xactRh}), and dashed lines are the large $m_0$ approximation Eq.~(\ref{proxRh}). Note optimal $\lambda_{\theta} \approx R^{\lambda_{\theta}}$. {\bf C.} Optimal $m_{\theta}/m_0$ as a function of $\rho$ and $m_0$.
    {\bf D.} Rate of food consumption $R^{\lambda_{\theta}^{H,L}}$ in binary depleting environments in which the observer knows patch type ($\lambda_H$ or $\lambda_L$) upon arrival. Optimal strategy (purple dot) takes $\lambda_{\theta}^H = \lambda_{\theta}^L \approx R^{\lambda_{\theta}^{H,L}} \approx 3.3$. $p_H = 0.5$, $m_{0}^H = 100$, $m_{0}^L = 50$, and $\rho = 0.1$. {\bf E.} Optimal waiting time $T_{\theta}^{\rm max}$ in environments where high (low) patch has one (zero) chunk(s) of food. {\bf F.} Optimal waiting time $T_{\theta}^{\rm max}$ and departure threshold $\lambda_{\theta}^{\rm max}$ in a binary environment where the low patch has zero chunks of food and the high patch initiall has $m_0^H = 50$. $\rho = 0.1$. Transit time $\tau = 5$ throughout.}
    \label{fig6:deplete}
\end{figure*}

We start by considering an agent foraging in an environment in which all patches begin with the same density of food, so the arrival rate in each is initially $\lambda_0$. As they forage, the arrival rate $\lambda (t) = \lambda_0 - K(t)\rho$ decreases in equal increments with each food encounter (Fig.~\ref{fig6:deplete}A). We assume the agent perfectly knows this arrival rate as well as the amount of time it has been foraging, and will use this information to determine a strategy for departing the patch. In keeping with the simplicity of the space of possible strategies discussed before, we will assume the forager departs the patch when $\lambda (t)$ (or analogously the density of food in the patch) falls below some threshold $\lambda_\theta$.

This problem can be framed by simply computing the mean food intake rate for a given strategy parameterized by $\lambda_\theta$. Assuming $\lambda_{\theta}$ is an integer multiple of $\rho$, we can compute directly the number of chunks consumed before departure $m_{\theta} \equiv K(T(\lambda_{\theta})) := (\lambda_0 - \lambda_{\theta})/\rho$, and the mean time to depart $\bar{T}(\lambda_{\theta})$. Linearity of expectations allows us to compute the mean departure time as the sum of mean exponential waiting times between food encounters
\begin{align*}
    \bar{T}(\lambda_{\theta}) &= \sum_{j=1}^{m_{\theta}} \frac{1}{\lambda_0 - (j-1)\rho} = \frac{1}{\rho} \sum_{j=1}^{m_{\theta}} \frac{1}{m_0 - (j-1)} \\
    &= \frac{H_{m_0} - H_{m_0 - m_{\theta}}}{\rho},
\end{align*}
where $H_n$ is the $n$th harmonic number. Thus, we compute the long term reward rate given $\lambda_{\theta}$ as
\begin{align}
    R^{\lambda_{\theta}} = \frac{m_0 \rho - \lambda_{\theta} }{H_{m_0} - H_{\lambda_{\theta}/\rho} + \rho \tau}. \label{xactRh}
\end{align}
There is an interior optimum $m_{\theta}$ that maximizes long term food consumption rate (Fig.~\ref{fig6:deplete}B,C). For low initial rate $\lambda_0$ and decrement $\rho$, the best strategy is to remain in a patch until all food is consumed, but as $\lambda_0$ and $\rho$ are increased, the optima occur roughly where $R^{\lambda_{\theta}} \equiv \lambda_{\theta}$ as in the marginal value theorem (MVT).

To analyze this observation further, we can approximate optima in the limit of plentiful patches ($m_0 \gg 1$), such that $H_n$ can be estimated by the large $n$ expansion
\begin{align}
    H_n = \log n + \gamma + {\mc O}(1/n), \label{harmasym}
\end{align}
where $\gamma = 0.57721566...$ (the Euler–Mascheroni constant).  Using this, a first approximation yields
\begin{align}
    R^{\lambda_{\theta}} \approx \frac{m_0 \rho - \lambda_{\theta} }{\log (\rho m_0) - \log \lambda_{\theta} + \rho \tau}. \label{proxRh}
\end{align}
In this limit, we can explicitly show the MVT is satisfied, $R^{\lambda_{\theta}} \equiv \lambda_{\theta}$, and the observer departs the patch when the the arrival rate of food in the patch matches their average environment-wide food arrival rate. To show this, note that $R^{\lambda_{\theta}}$ in Eq.~(\ref{proxRh}) has derivative
\begin{align*}
    \frac{d R^{\lambda_{\theta}}}{d \lambda_{\theta}} = \frac{-(\log (\rho m_0) - \log \lambda_{\theta} + \tau / \rho) + (m_0 \rho - \lambda_{\theta})/\lambda_{\theta}}{(\log (\rho m_0) - \log \lambda_{\theta} + \rho \tau )^2},
\end{align*}
which is increasing at $\lambda_{\theta} = \rho$:
\begin{align*}
    \left.\frac{d R^{\lambda_{\theta}}}{d \lambda_{\theta}}\right|_{\lambda_{\theta} = \rho} = \frac{(m_0-1)-(\log (m_0) + \tau / \rho)}{(\log m_0 + \rho \tau )^2} > 0,
\end{align*}
since $m_0 \gg 1$, and decreasing at $\lambda_{\theta} = \rho m_0 = \lambda_0$:
\begin{align*}
    \left. \frac{d R^{\lambda_{\theta}}}{d \lambda_{\theta}} \right|_{\lambda_{\theta} = \lambda_0} = - \frac{1}{\rho \tau} < 0.
\end{align*}
In between lies a critical point satisfying
\begin{align}
    \frac{m_0 \rho - \lambda_{\theta}}{\lambda_{\theta}} = \log \frac{\rho m_0}{\lambda_{\theta}} + \rho \tau, \label{mvtcp}
\end{align}
which is precisely the equation recovered when setting $R^{\lambda_{\theta}} \equiv \lambda_{\theta}$ in Eq.~(\ref{proxRh}), implying the critical point of $R^{\lambda_{\theta}}$ occurs where $R^{\lambda_{\theta}} \equiv \lambda_{\theta}$ as in the MVT.

Alternatively, for small $m_0$ (few chunks of food per patch), computing the harmonic numbers in Eq.~(\ref{xactRh}) is straightforward, so we can directly compare the mean rate $R^{\lambda_{\theta}}$ for all possible values of $\lambda_{\theta}$. For instance, when $m_0 = 2$, we can choose from $\lambda_{\theta} \in \{ 0, \rho \}$ as the departure threshold, corresponding to maximizing
\begin{align*}
    R^{\lambda_{\theta}} \in \left\{ \frac{4 \rho}{3 + 2\rho \tau}, \frac{2\rho}{1+2 \rho \tau} \right\} = \left\{ R^{\lambda_{\theta} =0}, R^{\lambda_{\theta} = \rho} \right\}.
\end{align*}
With this, the low threshold should be chosen when $ \rho > 1/(2 \tau)$, or when energy gained from food chunks is larger than half the transit rate between patches. At $\rho = 1/(2 \tau)$, the forager should leave after a single chunk is consumed, yielding $R^{\rm opt} = \rho = \lambda_{\theta}^{\rm opt}$ (as in the MVT). Similar results can be computed in the cases $m_0 = 3, 4, 5, ...$ using explicit computation of the harmonic numbers $H_n$.
%A similar result holds when $m_0 = 3$ where possible strategies are $\lambda_{\theta} \in \{ 0, \rho, 2 \rho \}$ so we maximize
%\begin{align*}
%    R^{\lambda_{\theta}} \in \left\{ \frac{18 \rho}{11 + 6\rho \tau}, \frac{12\rho}{5+6 \rho \tau}, \frac{3\rho}{1+3 \rho \tau} \right\},
%\end{align*}
%so the forager should clear the patch ($\lambda_{\theta} = 0$) if $\rho > 7/(6 \tau)$, eat till a chunk remains ($\lambda_{\theta} = \rho$) if $7/(6 \tau)>\rho>1/(6\tau)$, and eat a single chunk ($\lambda_{\theta} = 2\rho$) if $1/(6 \tau)>\rho$. As in the $m_0 = 2$ case, this is consistent with the MVT as at $\rho = 7/(6 \tau)$, $R^{\rm opt} = \rho = \lambda_{\theta}^{\rm opt}$ and at $\rho = 1/(6 \tau)$ we have $R^{\rm opt} = 2 \rho = \lambda_{\theta}^{\rm opt}$. 
We expect these results to generalize to arbitrary $m_0$ as evidenced by the small $m_0$ case and our large $m_0$ asymptotics.

Overall this suggests that for short transit times $\tau$ or small food chunk sizes $\rho$, and ideal observer will depart patches more quickly than when transit times are long $\tau$ or food chunks have a higher quality $\rho$. This is consistent with classic models of continuous consumption as in the MVT, but here we have provided a detailed analysis of the case of discretized and random food encounters. The primary reason for a mismatch with the MVT arises due to the discreteness of the food availability and arrival rate. We show in binary environments, when uncertainty plays a role in patch leaving decisions, there are nontrivial deviations from the MVT due to the forager not precisely knowing the food arrival rate of their current patch.

%\begin{itemize}
%\item Analyze in more detail patch leaving strategy in homogeneous environment, for maximizing long term energy yield.
%\begin{itemize}
 %   \item The optimal strategy in homogeneous environment with discrete rewards is (I think) 'counting', because you can exploit the information you have about the environment.  Show this theoretically?  
%    \item This differs from the MVT, which uses a rate-base rule to leave the patch.
%\end{itemize}
%\end{itemize}

\subsection{Binary environments}

Next, we consider binary depleting environments in which the observer must both infer the type of patch (high or low) and track its depletion. We focus on strategies in which the observer departs the patch when the mean estimate of the arrival rate $\tilde{\lambda}(t)$ falls below a threshold $\lambda_{\theta}$. To understand this most general case, we begin by discussing two simplified cases: (a) the case in which the patch type is known on arrival, and (b) the case in which the low yielding patch is empty ($m_0^L = 0$).

{\em Depletion-dominated regime.} One way of contextualizing the case of known patch types is that of depletion-dominated decisions. These types of decisions occur if the observer stays in a patch till they have determined the patch type to high certainty. To idealize this situation, assume the observer arrives in a patch and immediately knows the patch type $\lambda_j$ ($j \in \{ H, L\}$) in which they reside. Moreover, assume they leverage a strategy in which they depart the patch as soon as the arrival rate falls below some level $\lambda_{\theta}^j$. In this case, there are two thresholds to tune for each of the patch types. Following our calculations from the homogeneous case, the time to depart a patch of type $j$ is
\begin{align*}
\bar{T}_j(\lambda_{\theta}) = \frac{H_{m_0^j} - H_{m_0^j - m_{\theta}^j}}{\rho},
\end{align*}
where $m_0^j$ is the initial food count in patches of type $j$ and $m_{\theta}^j$ is the threshold level of food to consume before departing. As such, the long term reward rate is
\begin{align}
    R^{\lambda_{\theta}^{H,L}} = \frac{p_H(m_0^H \rho - \lambda_{\theta}^H) + p_L (m_0^L \rho - \lambda_{\theta}^L) }{p_H (H_{m_0^H} - H_{
    \lambda_{\theta}^H/\rho}) + p_L(H_{m_0^L} - H_{\lambda_{\theta}^L/\rho}) + \rho \tau }, \label{Rlambin}
\end{align}
maximized by setting $\lambda_{\theta}^H = \lambda_{\theta}^L \approx R^{\lambda_{\theta}^{H,L}}$ (Fig.~\ref{fig6:deplete}D). We show now that this follows the MVT.

In large $m_0^j$ limit, the harmonic numbers are approximated by Eq.~(\ref{harmasym}), and Eq.~(\ref{Rlambin}) becomes
\begin{align}
    R^{\lambda_{\theta}^{H,L}} \approx \frac{p_H(m_0^H \rho - \lambda_{\theta}^H) + p_L (m_0^L \rho - \lambda_{\theta}^L) }{p_H \tilde{T}_H(\lambda_{\theta}^H) + p_L \tilde{T}_L(\lambda_{\theta}^L) + \rho \tau },  \label{bindepknow}
\end{align}
where $\tilde{T}_j(\lambda_{\theta}^j) = \log (\rho m_0^j) - \log \lambda_{\theta}^j$ and the critical point equations for each partial derivative $R^{\lambda_{\theta}^{H,L}}_{\lambda_{\theta}^j} = 0$ imply
\begin{align}
& p_H (m_0^H - \lambda_{\theta}^H) + p_L (m_0^L \rho - \lambda_{\theta}^L) = \lambda_{\theta}^j \left[ p_H(\log (\rho m_0^H) \right. \nonumber \\
& \hspace{1cm} \left. - \log \lambda_{\theta}^H) + p_L (\log (\rho m_0^L) - \log \lambda_{\theta}^L) + \rho \tau \right], \label{bindepcp}
\end{align}
which can be rewritten as $R^{\lambda_{\theta}^{H,L}} = \lambda_{\theta}^j$ ($j = H,L$), implying the observer should depart a patch when the arrival rate equals the mean rate of food arrival for the environment (MVT). As in the homogeneous depleting environment, obtaining the prediction of the MVT relies on the observer knowing the current food arrival rate. However, as we will show, the optimal strategy deviates from the MVT when there is uncertainty concerning the present food arrival rate in the patch. 

{\em Empty low patch.} Before addressing the general case of a binary environment with initially unknown patch types, we study a simple case in which the forager does not know the patch type they are in ahead of time but where the statistics of patch departures are still explicitly calculable: when the low yielding patch is empty.

The simplest case is that in which some patches are empty ($m_0^L = 0$) and others have a single chunk of food ($m_0^H=1$), as introduced by~\cite{McNamara_Houston_1980} but not analyzed in detail. Clearly, when a forager encounters a chunk of food they should leave the patch, since there is no food left thereafter. How long should they wait until departing a patch if food has yet to be encountered? This amounts to optimizing the waiting time $T_{\theta}$ till departure (or equivalently setting a threshold mean estimated arrival rate $\lambda_{\theta}$ at which they depart). Each patch visit thus falls into one of three categories: (a) visits to empty (low) patches lasting time $T_{\theta}$; (b) visits to high patches resulting in no food lasting time $T_{\theta}$; and (c) visits to high patches resulting in one food chunk lasting time $t< T_{\theta}$.

To determine an optimal strategy, we need only compute statistics for case (c). Since the food arrival rate in a high patch is $\rho$, the probability of encountering food when waiting a time $T_{\theta}$ is $1-\pi_{\theta} = 1-\e^{- \rho T_{\theta}}$, the mean time to wait is
\begin{align*}
    \bar{T}_H^{\theta} = \frac{\rho \int_0^{T_{\theta}} t \e^{- \rho t} dt}{1 - \e^{- \rho T_{\theta}}} = \frac{1 - (\rho T_{\theta} + 1)\e^{-\rho T_{\theta}}}{\rho (1 - \e^{- \rho T_{\theta}})},
\end{align*}
and the rate of consumption of a given strategy is
\begin{align*}
    R^{\theta} = \frac{p_H (1-\pi_{\theta})}{p_H(\pi_{\theta}T_{\theta} + (1-\pi_{\theta}) \bar{T}_H^{\theta}) + p_LT_{\theta} + \tau},
\end{align*}
which we can maximize by finding the critical point of $R^{\theta}$ in $T_{\theta}$. As $p_H$ is increased and as $\rho$ is decreased, the optimal $T_{\theta} = T_{\theta}^{\rm max}$ increases (Fig.~\ref{fig6:deplete}E). When high yielding patches are more frequent (higher $p_H$), one should stay in a patch longer, since they are more likely to encounter food. When the rate of chunk discovery $\rho$ is smaller, one should expect to wait longer to encounter food in a high patch, so the wait time should be increased.

Note, the observer does not leave the initially empty patches immediately, so their strategy (even if optimally tuned) does not agree with the MVT due to the observer's uncertainty about each patch's yield rate.

\begin{figure*}
    \centering
    \includegraphics[width=17cm]{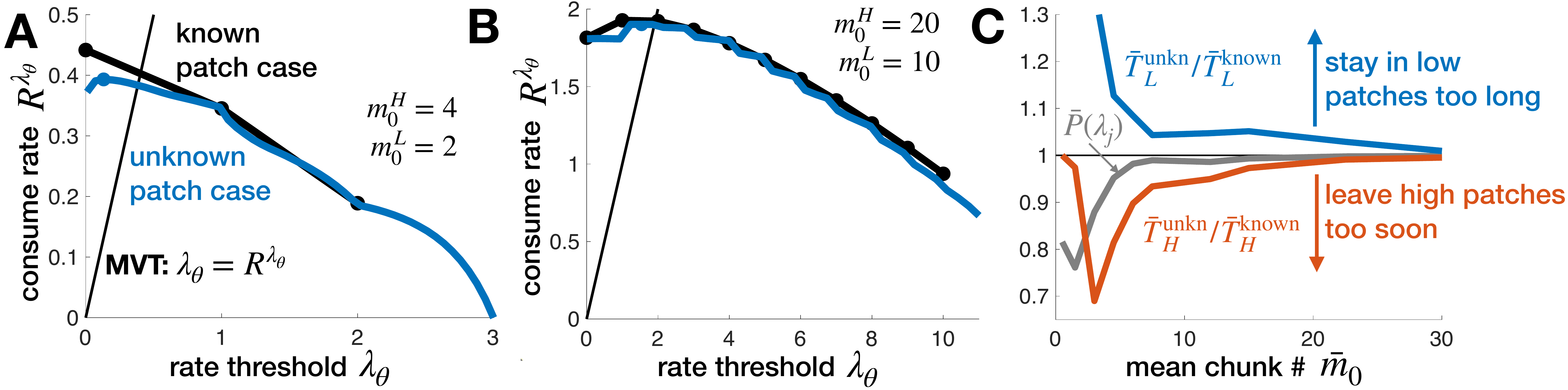}
    \caption{\textbf{Depleting environments with known versus unknown patch reward rates}.  In the `known' case, the forager knows the patch type and initial reward rate upon arrival, whereas in the `unknown' case the forager must infer the path type.
    Patch leaving strategies in the unknown case approach those of the known case in the high chunk count limit. {\bf A.} Total reward rate $R^{\lambda_{\theta}}$ as a function of the estimated arrival rate $\lambda_{\theta}$ at which the observer departs a patch. For low initial food levels $m_0^H$ and $m_0^L$, $R^{\lambda_{\theta}}$ near the optimum (dots) deviates between the case of initially known patch type (black) vs. initially unknown (blue). Both thresholds are less than the MVT prediction due to discreteness of chunk arrivals. {\bf B.} Reward rates of the known/unknown cases converge at higher initial food levels.
    {\bf C.} Deviations in optimal reward rate between known and unknown cases are due to less certainty $\bar{P}(\lambda_j)$ (grey line) about patch yield rate upon departure, leading the observer to stay in low patches too long (blue line) and leave high patches too soon (red line). Other parameters are $\rho = 1$, $\tau = 5$, and $p_H = 0.5$. Mean rates and departure times in the unknown patch case were computed from $10^6$ Monte Carlo simulations.}
    \label{fig7:bindepgen}
\end{figure*}

We also analyze the case of an empty low patch and an arbitrary number of food chunks $m_0$ in the high patch. Here, the observer should apply a hybrid strategy: Wait a finite time $T_{\theta}$ to depart if no food is encountered, but if food is encountered before $t=T_{\theta}$, consume food until the arrival rate drops to $\lambda_{\theta} = (m_0 - m_{\theta})\rho$. This strategy distinguishes two types of patch leaving decisions arising in uncertain and depleting environments. Decisions may be uncertainty-dominated, where the observer may depart a patch early, believing they are in a low-yielding patch. Alternatively, decisions may be depletion-dominated, after the observer is nearly certain of they patch type they are in and has depleted it.
%These situations arise in the more general case we consider below, but the case at hand is more tractable.

To optimally parameterize the departure strategy, we again compute the probability of departing the high patch early $\pi_{\theta} = \e^{- m_0 \rho T_{\theta}}$ (at time $T_{\theta}$). Assuming the observer remains in the patch, it takes a time
\begin{align*}
\bar{T}_H^{\theta} = \frac{1 - (m_0 \rho T_{\theta} + 1) \e^{- m_0 \rho T_{\theta}}}{m_0 \rho (1 - \e^{- m_0 \rho T_{\theta}})}    
\end{align*}
to consume the first chunk of food and then depart immediately ($T(\lambda_{\theta}) = 0$) if $\lambda_{\theta} = \lambda_0 - \rho$ (as in the previous case). Otherwise, it will take an additional time
\begin{align*}
    \bar{T}(\lambda_{\theta}) &= \sum_{j=2}^{m_{\theta}} \frac{1}{\lambda_0 - (j-1) \rho} = \frac{1}{\rho} \sum_{j=1}^{m_{\theta}-1} \frac{1}{m_0 - j} \\
    &= \frac{H_{m_0 - 1} - H_{m_0 - m_{\theta}}}{\rho}.
\end{align*}
At each patch then, the observer either obtains no food and departs after time $T_{\theta}$ or obtains $m_{\theta} = m_0- \lambda_{\theta} /\rho$ chunks and departs on average after time $\bar{T}_H^{\theta} + T(\lambda_{\theta})$. The probability of the latter is $p_H(1-\pi_{\theta})$,
%so
%\begin{align*}
%    R(T_{\theta}, \lambda_{\theta}) = \frac{p_H(1-\pi_{\theta}) (m_0 \rho - \lambda_{\theta}) }{p_H(1-\pi_{\theta})(H_{m_0 - 1} - H_{m_0 - m_{\theta}} + \bar{T}_H^{\theta}) + (1-p_H(1-\pi_{\theta})) \rho T_{\theta} + \rho \tau}.
%\end{align*}
and assuming $m_0 \gg 1$, we can approximate $H_{m_0 - 1} - H_{m_0 - m_{\theta}} \approx \log (m_0 - 1) -  \log (m_0 - m_{\theta})$, so the food consumption rate is approximated by the a continuous function
\begin{align*}
    R(T_{\theta}, \lambda_{\theta}) \approx \frac{p_H(1-\pi_{\theta}) (m_0 \rho - \lambda_{\theta}) }{p^{\theta} (\bar{T}(\lambda_{\theta}) + \bar{T}_H^{\theta}) + (1-p^{\theta}) \rho T_{\theta} + \rho \tau},
\end{align*}
where $p^{\theta} = p_H(1-\pi_{\theta})$ and $\bar{T}(\lambda_{\theta}) = \log (\rho m_0 - \rho) -  \log \lambda_{\theta}$.  Reward rate is maximized using a waiting time $T_{\theta}$ that is fairly insensitive to $p_H$, whereas the threshold $\lambda_{\theta}$ is sensitive to $p_H$ (Fig.~\ref{fig6:deplete}F). When high yielding patches become more common, the forager need not deplete them as much, since the next patch they visit is likely to be high yielding.

{\em Many food chunks.}
%We also consider the case in which $0 < \rho \ll 1$ and $m_0^j \gg 1$, so depletion of a patch is slow. In this limit, we expect the observer to usually discern the patch type long before they depart. To motivate this assumption, we refer to our results in the known patch type case (Fig.~\ref{fig6:deplete}D). Note that for $\rho$ small, most parameter regimes suggest the observer should wait until all food in patch has been consumed before transiting. 
Lastly, we examine general binary depleting environments in which $m_0^H > m_0^L$ are arbitrary integers, so the belief $y(t) = \log \frac{P(\lambda_H - K(t)\rho|x(t))}{P(\lambda_L - K(t)\rho|x(t))}$ evolves according to Eq.~(\ref{bindep}). In line with our previous analysis, the observer estimates the current arrival rate of the patch from this belief,
\begin{align}
\tilde{\lambda}(t) &=  \frac{\lambda_H + \e^{-y} \lambda_L}{1+ \e^{-y}} - \rho K(t), \label{lamtildep}
\end{align}
and departs the patch when $\tilde{\lambda}(t) \leq \lambda_{\theta}$. As before, we are concerned with tuning the threshold $\lambda_{\theta}$ so the long term reward rate
\begin{align*}
    R^{\lambda_{\theta}} = \frac{p_H \bar{m}_H + p_L \bar{m}_L}{p_H \bar{T}_H + p_L \bar{T}_L + \tau}
\end{align*}
is maximized. Unlike the non-depleting case, we cannot frame and explicitly solve a MFPT problem to determine the mean times spent $\bar{T}_H$ and $\bar{T}_L$ and number of food chunks consumed $\bar{m}_H$ and $\bar{m}_L$ in the high and low patches. However, we can compute these quantities via Monte Carlo sampling. There is an internal optimum $\lambda_{\theta}^{\rm opt}$ that maximizes the reward rate $R^{\lambda_{\theta}}$ across the environment, which deviates only slightly when little food is available (Fig.~\ref{fig7:bindepgen}A) and is well matched in the case of high food availability (Fig.~\ref{fig7:bindepgen}B) to a forager that knows the patch type upon patch arrival (compare black/blue curves). This suggests the forager typically learns the patch type before departing in environments with sufficient food, so their departure strategy will converge to that of an observer who already knows the patch type.

We analyzed this trend by comparing departure times of observers that initially know their current patch type to those that do not (Fig.~\ref{fig7:bindepgen}C). Indeed, foragers in sparser environments (lower average initial food amount $\bar{m}_0 = (\bar{m}_0^H + \bar{m}_0^L)/2$) stay in low patches too long (blue curve) and leave high patches too soon (red curve) in comparison to observers that immediately know their patch type. This is due to their uncertainty about their current patch before and at the time of departure (grey curve). On the other hand, in more plentiful environments, foragers accumulate enough evidence about the current patch that their departure times resemble those of observers that know their patch type, recapitulating the MVT.

Uncertainty thus drives deviations from the MVT. Foragers that exit patches before fully learning their type will typically understay (overstay) high (low) patches. Key to these uncertainty-driven decisions is an environment in which high/low patches are different enough to warrant different strategies, but similar enough as to not be immediately distinguishable. We now extend our analysis to the case of environments small enough so that foragers eventually return to previously depleted patches.

%(Note for discussion:  when have continuous rewards, and therefore have perfect estimate of current rate of return, the prior doesn't matter - this is the conditions for the MVT. However, when rewards arrive discretely, the prior  matters)

\begin{figure*}
    \centering
    \includegraphics[width=16cm]{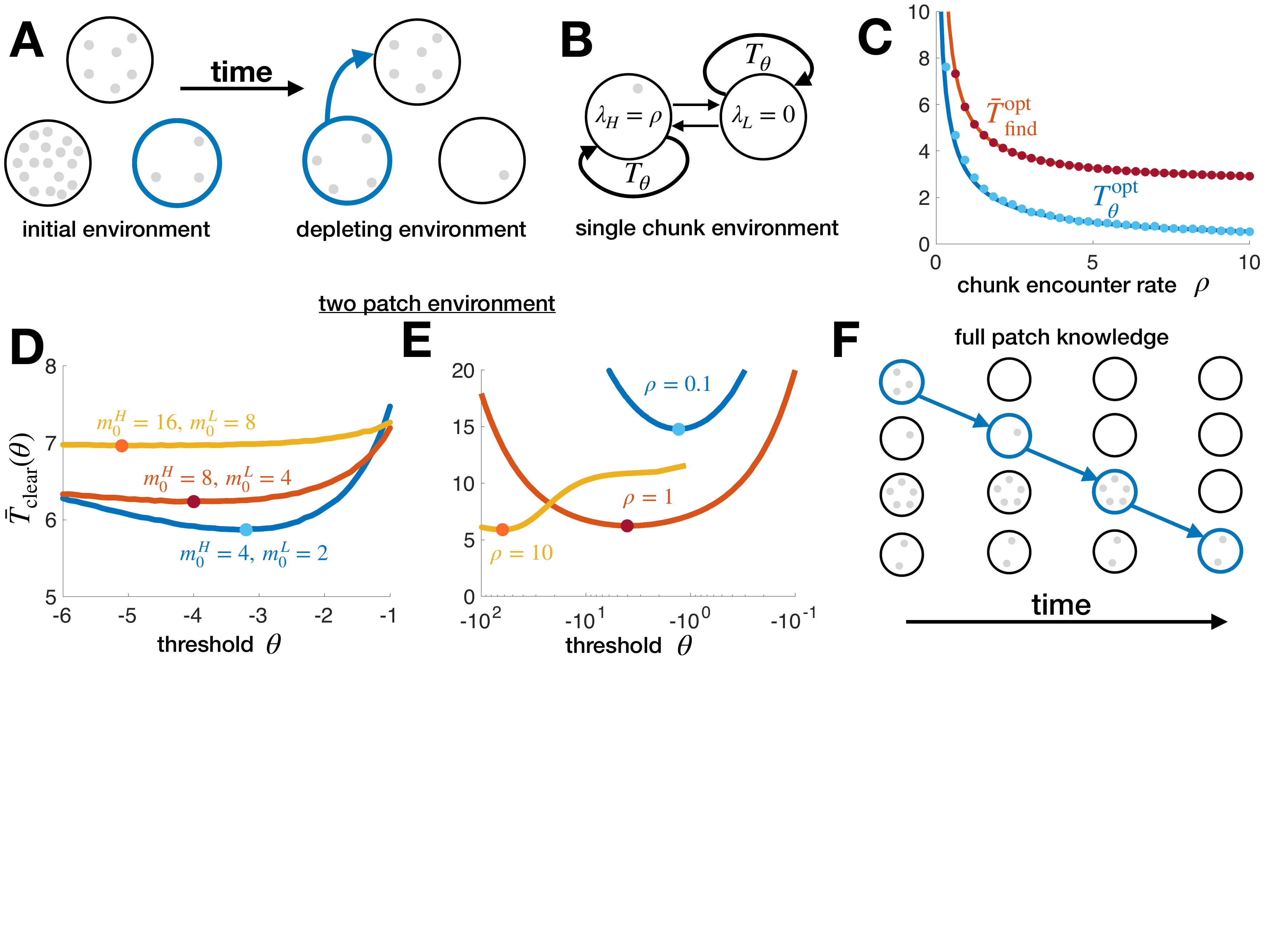}
    \caption{\textbf{Returning to patches in fully depleting environments}. {\bf A.} Environmental food depletion in small environments where previously visited patches (blue) remain depleted indefinitely. {\bf B.} Strategy in a two patch environment with a single food chunk: Switch patches after a finite search time $T_{\theta}$. {\bf C.} Optimal switch time $T_{\theta}^{\rm opt}$ (approximated by Eq.~(\ref{topretform})) and the total mean time to find a chunk $\bar{T}_{\rm find}^{\rm opt}$ (Eq.~(\ref{tfinddepenv})) decrease as the rate of chunk detection $\rho$ increases. Theory compares well with averages from $10^6$ Monte Carlo simulations (dots). {\bf D.} In binary environments, the optimal LLR threshold $\theta^{\rm opt}$ (dots) for minimizing the time to clear the environment decreases as the amount of food in the environment increases. $\rho =1$ is used here. {\bf E.} Optimal threshold $\theta^{\rm opt}$ decreases as the rate of encountering a chunk increases: $m_0^H = 8$, $m_0^L = 4$. $p_H = 0.5$ and $10^6$ Monte Carlo simulations for each curve. {\bf F.} Optimal strategy given full knowledge of chunk count in each patch is to clear each patch sequentially.}
    \label{fig8:depenv}
\end{figure*}

\subsection{Returning to a previously depleted patch}

Assuming foragers do not return to previously depleted patches is a reasonable approximation in large environments, since there is a low probability of the observer returning to a previously visited patch if drawing randomly from their environment. However, in small environments, past patch depletion impacts the value of future patch visits sooner (Fig.~\ref{fig8:depenv}A). We briefly analyze this case now, identifying strategies that minimize the time to fully deplete the environment before departing for another environment.

{\em Single chunk-two patch environment.} First, consider the simplest nontrivial case, a two patch environment in which there is a single chunk of food ($m_0^H = 1$ so $\lambda_H = \rho$ initially and $m_0^L = 0$). When the forager departs one patch, they journey to the other patch with transit time $\tau$. We seek a waiting time strategy similar to that developed for depleting patches (Fig.~\ref{fig8:depenv}B). The observer waits a time $T_{\theta}$ in their current patch before leaving to the other patch (if they do not encounter a food chunk). Note, this differs from thresholding the LLR ($y = \theta$) since the first patch visit would be shorter in that case, but this approach allows an explicit approximation of the optimal strategy. If the forager is in the high patch, they encounter the food chunk with probability $1- \pi_{\theta} = 1 - \e^{- \lambda_H T_{\theta}}$. In the low patch, they always depart after a time $T_{\theta}$. Thus, the mean time to encounter the chunk in the environment is given by: (a) the time spent in the high patch visit on which the chunk is discovered plus (b) the initial time spent in the low patch when that is the first patch visited (scaled by that probability 1/2) plus (c) the other time spent visiting patches and not finding the chunk:
\begin{align}
   &\bar{T}_{\rm find} (T_{\theta}; \lambda_H) =  \frac{1 - (\lambda_H T_{\theta} + 1) \e^{- \lambda_H T_{\theta}}}{\lambda_H ( 1- \e^{- \lambda_H T_{\theta}})} + \frac{T_{\theta} + \tau}{2} \nonumber \\
   & \hspace{3cm} + 2 (T_{\theta} + \tau) (1- \pi_{\theta}) \sum_{j=0}^{\infty} j \pi_{\theta}^j \nonumber \\
   & \hspace{5mm} = \frac{\lambda_H T_{\theta} + 3 \lambda_H \tau - 2 + (2 + \lambda_H T_{\theta} + \lambda_H \tau)\e^{\lambda_H T_{\theta}}}{2 \lambda_H (\e^{\lambda_H T_{\theta}} - 1)}. \label{tfinddepenv}
\end{align}
This is minimized by taking $T_{\theta} = T_{\theta}^{\rm opt}$ such that
\begin{align*}
    \e^{2 \lambda_H T_{\theta}^{\rm opt}} - 2(\lambda_H T_{\theta}^{\rm opt} + 2\lambda_H \tau) \e^{\lambda_H T_{\theta}^{\rm opt}} - 1 = 0,
\end{align*}
For large $\e^{\lambda_H T_{\theta}^{\rm opt}}$, we approximate the critical point equation as $\e^{\lambda_H T_{\theta}^{\rm opt}} \approx 2 \lambda_H T_{\theta}^{\rm opt} + 4 \lambda_H \tau$, which has a compact asymptotic expansion:
\begin{align}
   T_{\theta}^{\rm opt} & \approx \frac{1}{2 \lambda_H} \left[ - 2 W_{-1} \left( - \frac{\e^{- 2\lambda_H \tau}}{2} \right) - 4 \lambda_H \tau \right] \nonumber \\
   & \approx \frac{1}{\lambda_H} \log (4 \lambda_H \tau + 2\log 2), \label{topretform}
\end{align}
where $W_{-1}$ is the $-1$ branch of the Lambert W function.  This shows that the optimal waiting time scales inversely with the initial arrival rate $\lambda_H$ of the high patch. The minimal time to find the chunk is approximated by substituting Eq.~(\ref{topretform}) into Eq.~(\ref{tfinddepenv}).
%\begin{align*}
%    \bar{T}_{\rm find} (T_{\theta}^{\rm opt}; \lambda_H) \approx \frac{\log (g(\lambda_H, \tau)) + 3 \lambda_H \tau - 2 + (2 + \log (g(\lambda_H, \tau)) + \lambda_H \tau)g (\lambda_H, \tau)}{2 \lambda_H (g(\lambda_H, \tau) -1)},
%\end{align*}
%where $g(\lambda_H, \tau) = 4 \lambda_H \tau + 2 \log 2$.
Both quantities decrease with $\rho$ (Fig.~\ref{fig8:depenv}C), owing to the fact that environments with higher discovery rates $\lambda_H = \rho$ require less time in the high patch for the observer to find the food chunk. When the detection rate $\rho$ is small, the optimal switch time $T_{\theta}^{\rm opt}$ (i.e. the decision strategy) is more sensitive to small changes in $\rho$.

{\em General binary environment.} A similar approach can be used in binary environments with an arbitrary initial number $m_0^H > m_0^L$ of food chunks in the high and low patch. To minimize the number of required patch switches, the forager should first clear $m_0^L$ chunks of food from the starting patch and then wait until the observer's LLR $y = \log \frac{P(\lambda_H(t)|x(t)))}{P(\lambda_L(t)|x(t))}$ (computed from  Eq.~(\ref{bindep})) falls below the threshold $\theta < \log \frac{p_H}{1-p_H}$. After the first couple of patch switches, the time spent in a patch if food is not discovered will be $T_{\theta} = - 2\theta/[\rho (m_0^H - m_0^L)]$.
%We show that the time $\bar{T}_{\rm clear} (\theta)$ to clear the environment of food will be a nonmonotonic function of the threshold $\theta$.
While an explicit formula for $\bar{T}_{\rm clear}(\theta)$ can be obtained by marginalization, the resulting expressions are lengthy and do not provide insight into the trends underlying the optimal strategy. Thus, we use Monte Carlo sampling to determine how the optimal $\theta$ varies with the environmental parameters.

As the amount of food in the environment increases, the time to clear it increases, but only slightly (Fig.~\ref{fig8:depenv}D). Since the observer has more evidence about the kind of patch they are in after consuming $m_0^L$ chunks as $m_0^L$ grows, they can more quickly find the $(m_0^L+1)$th chunk. Moreover, the patches become more discriminable as the difference $m_0^H - m_0^L$ increases, such that the forager can afford more certainty before switching patches, meaning that the optimal $\theta^{\rm opt}$ is lower. Increased discriminability also results from increasing the rate of chunk encounters $\rho$ (Fig.~\ref{fig8:depenv}E).  With increased discriminability, both the optimal threshold $\theta^{\rm opt}$ and mean time to clear the environment $\bar{T}_{\rm clear}^{\rm opt}$ decrease owing to the forager's ability to discover chunks more quickly.

{\em More patches.} We expect it is possible to extend the above strategy to the case of $N$ environmental patches. However, the dimensionality of the optimization problem will grow quickly, so we save a detailed analysis of this case for future work.

Nonetheless, we can easily identify a strategy for minimizing the time to clear the environment in the case of $N$ patches assuming the observer knows the patch type upon arrival. If patch $j$ has $m_0^j$ chunks of food initially, the forager should still aim to minimize the total time spent in transit between patches. To do this, they should fully clear each patch before moving on to the next (Fig.~\ref{fig8:depenv}F). Assuming the forager always chooses an unexplored patch when departing an emptied one, the minimal mean time to clear the environment of food will be
\begin{align*}
    \bar{T}_{\rm clear}^{\rm opt} = (N-1)\tau + \frac{1}{\rho} \sum_{j=1}^N H_{m_0^j} = N \left[ \tau + \bar{H}/ \rho \right] - \tau,
\end{align*}
where the time to clear the $j$th patch is $H_{m_0^j}/\rho$ as in Eq.~(\ref{xactRh}) for homogeneous depleting environments and $\bar{H}/\rho$ is the mean clearance time of patches across the environment. Thus, the time to clear the environment scales linearly with the number of patches $N$ (assuming the mean patch clearance time is unchanged by $N$). 

Were the forager to use another strategy, they would return to each patch multiple times to clear them. This would mean the transit contribution would be $\bar{T}_{\rm transit} > (N-1) \tau$, and the total time to clear patch $j$ would be no less than $H_{m_0^j}/\rho$, so $\bar{T}_{\rm clear} > \bar{T}_{\rm clear}^{\rm opt}$.

There are a number of extensions to this case of a fully depleting environment. Since the observer must consider both transit times between patches and patch yields that depend on visit history, this is a qualitatively different problem from the multi-armed bandit (MAB) problem. Typically, bandit problems do not involve depletion or action-dependent changes in the yield of arms~\cite{scott2010modern,banks_switching_1994}. Our setup thus provides a rich class of optimal switching strategy problems which warrant future investigation.

\section{Learning the distribution of patch types}
\label{sec:learn}

Our analysis so far has assumed the forager has complete knowledge of the distribution of patch arrival rates in the environment. However, typically animals arrive in a new environment with uncertainty about the quality of patches they will encounter~\cite{kamil1982learning,regelmann1986learning,Eliassen_Jorgensen_Mangel_Giske_2009,constantino2015learning}. Thus, animals should use information from patch encounters to estimate statistics of patch yields in the environment. In our sequential updating model, this can be implemented with a higher level of inference in the model (Fig.~\ref{fig1:schematic}B) whereby the observer refines a prior over the initial arrival rate $\lambda_0$ using the past $N$ patch visits $p(\lambda_0 | x_{1:N}(t))$. We assume each new patch a forager enters has not been previously depleted.

Here, we measure learning performance using the time it takes to refine the mean squared error (MSE) to some threshold. Interestingly, we find that foragers in depleting environments learn the initial rate of arrival faster than those in non-depleting environments.

\subsection{Homogeneous environment}

A homogeneous environment is described by a single parameter $\lambda_0$, the initial arrival rate of food within all patches. We will start by considering the simple non-depletion case, and then move to consider how learning occurs in depleting environments.  This will illuminate the differences between learning in these two cases before moving to binary environments.

{\em Non-depleting environments.} In a non-depleting environment, an agent intending to maximize the speed they learn the patch arrival rate of food $\lambda_0$ should remain in a single patch indefinitely. In this case, the rate learning process is equivalent to learning the rate of a sequence of exponentially distributed waiting times. Given a sequence of food encounters $x(t) = \sum_{j=1}^{K(t)} \delta (t - t_j)$, the observer uses the time $t$, count $K(t)$, and initial prior $p_0(\lambda_0)$ to estimate the rate $\lambda_0$ so Bayes rule states
\begin{align}
    p(\lambda_0|x(t)) &= \frac{1}{p(K(t))} p(K(t)|\lambda_0) p_0(\lambda_0) \nonumber \\
    & \propto (\lambda_0 t)^{K(t)} \e^{- \lambda_0 t} p_0(\lambda_0),  \label{nondeplearn}
\end{align}
and the maximum likelihood estimate (MLE) of the rate $\lambda_0^*$ is given by the implicit equation
\begin{align*}
    t \lambda_0^* &=  K(t) + \lambda_0^* p_0'(\lambda_0^*)/p_0(\lambda_0^*).
\end{align*}
For a flat, improper prior $p_0(\lambda_0) \equiv 1$, the MLE is $\lambda_0^* = K(t)/t$, and for an exponential prior  $p_0(\lambda_0) = a \e^{-a \lambda_0}$, the MLE is $\lambda_0^* = K(t)/(t+a)$. In either case, the MLE is consistent in the limit as $t \to \infty$:
\begin{align*}
    \lim_{t \to \infty} \lambda_0^* = \lim_{t \to \infty} \frac{K(t)}{t} +  \lim_{t \to \infty} \frac{\lambda_0^* p_0'(\lambda_0^*)/p_0(\lambda_0^*)}{t} = \lambda_0^{\rm true}.
\end{align*}

\begin{figure*}
    \centering
    \includegraphics[width=17cm]{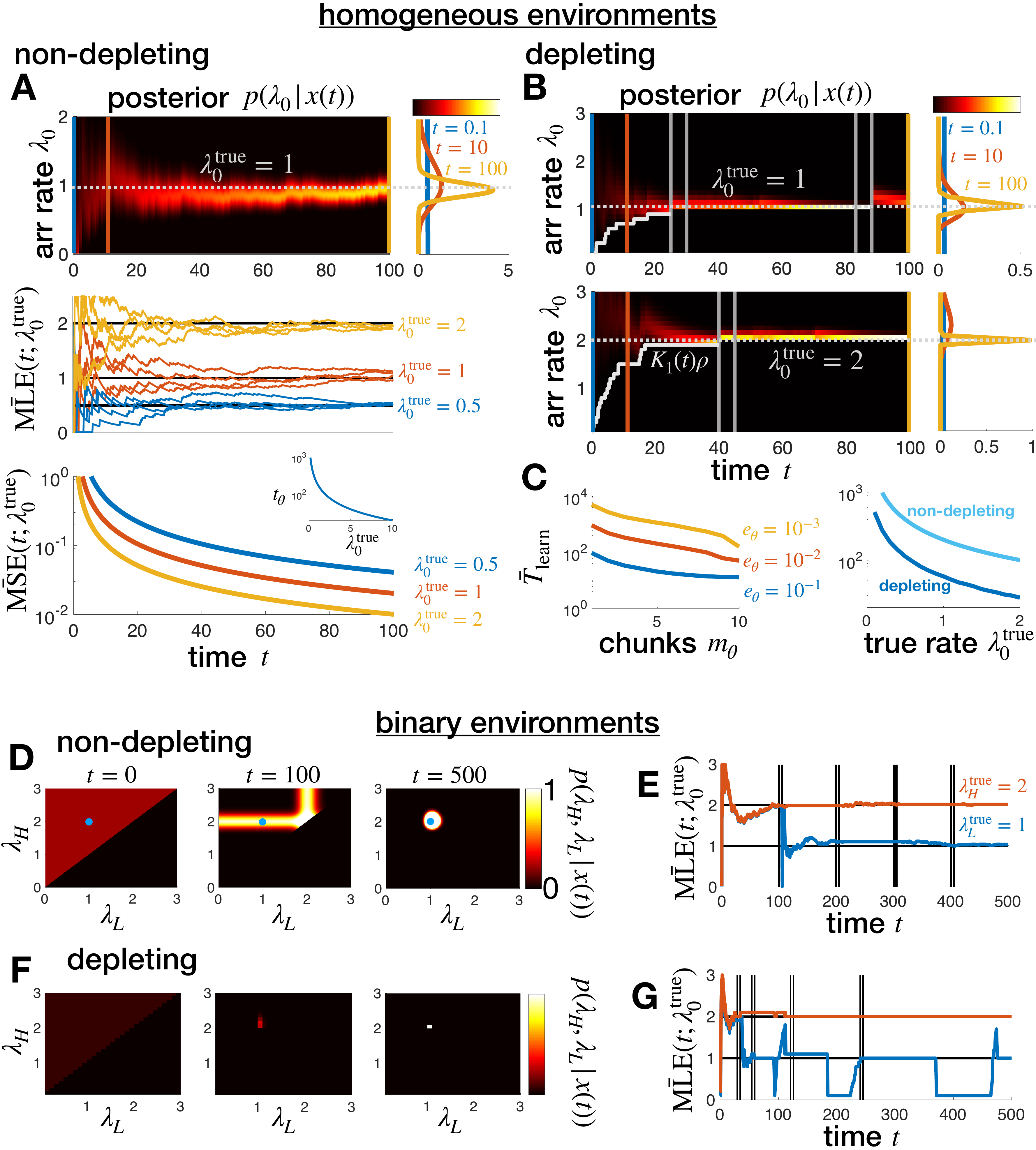}
    \caption{\textbf{Learning reward arrival rates}.  Patch arrival rates are learned more rapidly in depleting than non-depleting environments. {\bf A.} Posterior pdf $p(\lambda_0 | x(t))$ is refined leading to a maximum likelihood estimate (MLE) which approaches the true arrival rate: results for a non-depleting environment. Average mean squared error (MSE), given by Eq.~(\ref{msevthomno}), decreases more rapidly for higher $\lambda_0^{\rm true}$. {\bf B.} Posterior pmf $p(\lambda_0 | x(t))$ peaks more rapidly in depleting environments due to exclusion of rates $\lambda_0 < K(t) \rho$. {\bf C.}~Mean time to learn $\bar{T}_{\rm learn}$ to within MSE$<e_{\theta}$ is smallest when chunks consumed equals the total number in the patch $m_{\theta} = m_0 = 10$ (left), so the best policy (right) is to exit when MLE of total number of chunks is consumed ($m_{\theta}^{\rm opt} = {\rm MLE}(m_0)$). Here $
    \rho = 0.1$. Note lower $\bar{T}_{\rm learn}$ than in non-depleting case. {\bf D.} In binary environments, posterior pdf $p(\lambda_H, \lambda_L | x(t))$ is first refined over the first patch type visited ($\lambda_H = 2$ here), then over the second patch type visited ($\lambda_L = 1$), and eventually reaches a peak near true $(\lambda_H, \lambda_L)$. {\bf E.} Thus, eventually an accurate MLE is obtained for both patch types. Departures every $t=100$ time units. {\bf F,G.} Learning is more rapid in depleting environments, because low arrival rates are ruled out. Patches departed after MLE chunks consumed.}
    \label{fig9:learning}
\end{figure*}

Each food encounter tends to narrow the distribution $p(\lambda_0 | x(t))$ and the MLE tends to move toward the true arrival rate $\lambda_0^{\rm true}$ (Fig.~\ref{fig9:learning}A). One measure of learning is the rate of decrease of the normalized MSE. Since the estimator is consistent, we expect
\begin{align*}
    {\rm MSE}(t;\lambda_0^{\rm true}) = \int_0^{\infty} \frac{(\lambda_0 - \lambda_0^{\rm true})^2}{(\lambda_0^{\rm true})^2} p(\lambda_0 | x(t)) d \lambda_0
\end{align*}
to decrease on average over time. That is, $\frac{d}{dt} \bar{\rm MSE}(\lambda_0^{\rm true}, t) < 0$ where
\begin{align}
    \bar{\rm MSE}(t; \lambda_0^{\rm true}) = \int_0^{\infty} \frac{(\lambda_0 - \lambda_0^{\rm true})^2}{(\lambda_0^{\rm true})^2} \bar{p}(\lambda_0 | x(t)) d \lambda_0, \label{msevthomno}
\end{align}
and where the average is over realizations of arrival times $p(K(t)|\lambda_0^{\rm true})$, leading to 
\begin{align*}
    \bar{p}(\lambda_0 | x(t)) = \sum_{K=0}^{\infty} \frac{(\lambda_0^{\rm true}t)^K (\lambda_0 t)^K \e^{- \lambda_0^{\rm true} t} \e^{- \lambda_0 t} p_0(\lambda_0)}{(K!)^2 p(K)}.
\end{align*}
For a flat, improper prior $p_0(\lambda_0) \equiv 1$,
\begin{align*}
     p(\lambda_0 | K(t)) = \frac{t}{K!} (\lambda_0^t)^K \e^{- \lambda_0t}, 
\end{align*}
so
\begin{align}
    \bar{p}(\lambda_0 | x(t)) &= t \e^{- (\lambda_0^{\rm true} + \lambda_0) t} \sum_{K=0}^{\infty} \frac{(\lambda_0^{\rm true} \lambda_0 t^2)^K }{(K!)^2} \nonumber \\
    &= t I_0(2t \sqrt{\lambda_0 \lambda_0^{\rm true}})  \e^{- (\lambda_0^{\rm true} + \lambda_0) t} \label{fpmeanmsehomno} \\
    & \equiv F(t,\lambda_0, \lambda_0^{\rm true}), \nonumber
\end{align}
where $I_0(z)$ is the zeroth order modified Bessel function of the first kind. Eq.~(\ref{fpmeanmsehomno}) can thus be substituted into Eq.~(\ref{msevthomno}) and integrated using quadrature to calculate the ensemble mean of the MSE as a function of time:
\begin{align*}
    \bar{\rm MSE}(t;\lambda_0^{\rm true}) &= \int_0^{\infty} \frac{(\lambda_0 - \lambda_0^{\rm true})^2}{(\lambda_0^{\rm true})^2}  F(t,\lambda_0, \lambda_0^{\rm true}) d \lambda_0 \\
    & = \int_0^{\infty} (x - 1)^2  F(s,x, 1) d x,
\end{align*}
where we have made the changes of variables $\lambda_0^{\rm true} x = \lambda_0$ and $s = \lambda_0^{\rm true} t$. This reveals that the rate of decay of the normalized mean MSE increases linearly in $\lambda_0^{\rm true}$, such that the time $t_{\theta}$ to obtain a threshold level of mean accuracy scales inversely with $\lambda_0^{\rm true}$ (Fig.~\ref{fig9:learning}A bottom inset). Thus, the observer learns the arrival rate of the environment more quickly in higher rate environments.

Note, there are a number of ways to modify this basic setup. First, the forager may have a more informed prior over $\lambda_0$ based on experience with previous environments.  In this case the updating process is the same and only the prior $p_0(\lambda_0)$ changes. Had we considered an observer that moves between patches, the only difference would be that the observer gains no information about the arrival rate while in transit but carries their posterior forward upon arrival in the new patch. We now extend this framework to consider depleting environments.

{\em Depleting environments.} In depleting environments, the forager's belief update changes in two key ways. First, to continue learning, they must eventually depart their current patch and continue observing in a new patch whose arrival rate is reset to the true initial arrival rate $\lambda_0$. Second, each chunk encounter reduces the arrival rate within the patch, so the space of possible arrival rates must be adjusted. In particular, all arrival rates $\lambda_0 < K(t) \rho$ are eliminated from the posterior, as they are impossible given $K(t)$ chunks have arrived. With these facts in mind, we examine how the time to learn the arrival rate depends on patch exit strategy and true arrival rate. Following depleting foraging strategies, we assume the departure threshold on the number of chunks consumed depends on the MLE of the arrival rate. 

To determine an efficient strategy for inferring the arrival rate $\lambda_0$, we first determine the optimal departure chunk count given a specific true $\lambda_0^{\rm true}$. We then examine how an observer who chooses this chunk count threshold based on their MLE limits their time to obtain a given desired MSE. As before, the posterior over the initial arrival rate $\lambda_0$ can be calculated using Bayes rule, so within a single patch
\begin{align*}
    p(\lambda_0 | x(t)) & = \frac{1}{p(K(t))} p(K(t)|\lambda_0) p_0(\lambda_0) \propto \frac{m_0! \e^{- \lambda_0 t} p_0(\lambda_0)}{(m_0 - K(t))!} ,
\end{align*}
for $\lambda_0 \geq K(t) \rho$, where $m_0 = \lambda_0 / \rho$ is the initial count associated with the initial rate $\lambda_0$. Each food encounter generally narrows the distribution $p(\lambda_0 |x(t))$ and decreases the number of chunks in the patch. Given $N$ patch visits, the forager can accumulate evidence from each visit as a separate term in the posterior product:
\begin{align}
    p(\lambda_0 | x(t)) \propto &  \prod_{j=1}^N \frac{m_0! \e^{- \lambda_0  t_j}}{(m_0 - K_j(t_j))!} \cdot p_0(\lambda_0), \label{homlearndepj}
\end{align}
where $t_j$ is the time spent in the $j$th patch and $K_j(t_j)$ is the number of chunks encountered (Fig.~\ref{fig9:learning}B top).
%Assuming we wish the MSE about the true arrival rate to fall to some threshold level $e_{\theta}$, there will be an optimal number of chunks to consume before departing a patch to minimize the time to reach this threshold level.

Averages from simulations show the best strategy for minimizing the time to learn the true rate $\lambda_0^{\rm true}$ is to depart the patch when all the chunks predicted by the observer's current MLE are consumed (Fig.~\ref{fig9:learning}C left). We see evidence of this, for instance, when attempting to infer $\lambda_0^{\rm true} = 1$ (when $\rho = 0.1$ and $m_0 = 10$). The forager's best strategy for minimizing the time to learn this rate is to consume all 10 chunks of food in a patch before departing. Thus, an effective strategy for rapidly learning the initial arrival rate $\lambda_0$ across environments is to leave a patch once the number of chunks consumed is greater than or equal to the MLE for the initial chunk count $m_0$ ($m_{\theta}^{\rm opt} = {\rm MLE}(m_0)$). Using this strategy, the time to learn the rate decreases as the true rate increases (Fig.~\ref{fig9:learning}C right) in line with findings in the non-depleting case, but much faster due to exclusion effects of depletion.

\subsection{Binary environments}

A forager learning two patch arrival rates $\lambda_H>\lambda_L$ clearly must visit more than one patch, and each successive patch visit increases the likelihood that both patch types have been visited. As before, we start by considering the non-depleting environment in which learning is slower, and then move to studying the inference process in a depleting environment.

{\em Non-depleting environments.} In a single patch ($j$), given a flat prior ($p(\lambda_0)$ constant), a sequence of food encounters $x_j(t)$ yields the posterior
\begin{align*}
    p(\lambda_0 | x_j) \propto (\lambda_0 t_j)^{K_j(t_j)} \e^{- \lambda_0 t_j} \equiv Q(\lambda_0; t_j, K_j),
\end{align*}
over the possible arrival rate $\lambda_0$ in patch $j$, where $t_j$ is the time spent and $K_j(t_j)$ is the number of chunks encountered in patch $j$ (as in Eq.~(\ref{nondeplearn})).

Given $N$ patch visits and a flat prior ($p(\lambda_H, \lambda_L)$ constant) over the rates, the forager combines information across patches by conditioning on the probability they are in a high/low patch. The posterior for the arrival rates $\lambda_k$ ($k = H,L$) given patch visits $j=1,...,N$ is then:
\begin{align*}
    p(\lambda_H, \lambda_L |x(t)) &\propto \prod_{j=1}^N p(x_j(t_j)|\lambda_H, \lambda_L), \ \ \ \ \lambda_H > \lambda_L, \\
    & \propto \prod_{j=1}^N \left[ p_H Q(\lambda_H; t_j, K_j) + p_L Q(\lambda_L; t_j, K_j) \right],
\end{align*}
where $p_k \equiv p(\lambda_j = \lambda_k)$ is assumed known to the observer. 

The posterior updating is thus refined based on the arrival rate of the current patch (about $\lambda_H^{\rm true}$ when in a high patch and about $\lambda_L^{\rm true}$ when in a low patch: Fig.~\ref{fig9:learning}D). This process continues as the agent visits each type of patch, as seen in time series of the MLE, whereby the estimate of the arrival rate for the patch not being visited remains fairly constant, while that of the visited patch changes (Fig.~\ref{fig9:learning}E).

{\em Depleting environments.} Lastly, we derive the inference strategy for learning in binary environments with depleting patches. Learning the arrival rate of a single patch is given by Eq.~(\ref{homlearndepj}), and using a flat prior we have
\begin{align*}
    p(\lambda_0|x_j(t)) \propto \frac{(\lambda_0/\rho)!}{(\lambda_0/\rho-K_j)!} \e^{-\lambda_0 t_j} \equiv Q_{\rho}^j (\lambda_0),
\end{align*}
where $t_j$ is time spent and $K_j$ is the number of chunks encountered in patch $j$. Conditioning on the patch type being visited yields the posterior for the initial arrival rates $\lambda_0^k$ ($k=H,L$) given patch visits $j=1,2,...,N$:
\begin{align*}
    p(\lambda_0^H, \lambda_0^L | x(t)) &\propto \prod_{j=1}^N p(x_j(t_j)|\lambda_0^H, \lambda_0^L), \ \ \ \ \lambda_0^H > \lambda_0^L, \\
    & \propto \prod_{j=1}^N \left[ p_H Q_{\rho}^j(\lambda_H) + p_L Q_{\rho}^j(\lambda_L) \right],
\end{align*}
where $p_k \equiv p(\lambda_0^j = \lambda_0^k)$.

As in the non-depleting environment, the posterior is first refined over the first patch type visited ($\lambda_H = 2$ in Fig.~\ref{fig9:learning}F). However, as opposed to the homogeneous environment, low rates $\lambda_L < K \rho$ are never entirely ruled out, since there is always a possibility the forager is in the high patch. As such, the MLE for the high patch is refined more quickly while the low patch MLE continues to jump around (Fig.~\ref{fig9:learning}G).

We could also consider models that learn the depleting increment $\rho$ or distribution of transit times $\tau$ using a similar Bayesian sequential updating framework. Presumably, natural foragers would learn environmental parameters through related evidence accumulation processes. Our work here demonstrates that sequential Bayesian inference affords a powerful framework for understanding how a forager could learn hierarchically using a fairly compact model. A key finding of the above analysis is that depletion itself provides additional information about the arrival rate, allowing the observer to refine their estimate more rapidly than in non-depleting environments.

%\begin{itemize}
    %\item Homogeneous environment with depletion, just learn a single initial arrival rate $\lambda_0$
    %\begin{itemize}
     %   \item Could consider prior for distribution of $\lambda_0$.  Others have looked at similar problem in more detail.  Here, treat a simple/reasonable case (don't treat all/general).
    %\end{itemize}
    %\item Or two patch types with no depletion: Have to learn a joint posterior over $\lambda_H$ and $\lambda_L$
    %\item Focus on optimal strategies for most rapidly learning the arrival rates.
    %\item learning in a spatially structured environment. how does learning changes with the spatial arrangement of the patches. How to introduce particular search / planning computations into the behavior. 
    %the issue of spatial learning comes up frequently when one is talking about spatially inhomogeneous environment that requires a sort of planning or search which many would assume will be better described via a type of RL. I know I am pushing this a lot but I get this question a lot from the cognitive neuroscience crowd. Having criteria by which one can delineate a quantitative manner by which one can compare an RL mediated versus Bayesian inference mediated mechanism. In the same manner how would one get at a hybrid frame where the bayesian inference machinery is used to approximate or predict the state of the environment afterwards the RL machinery takes over.. 
%\end{itemize}

%one would hope to get at something like a learning rate curve and its dependence on the environmental parameters. 

\section{Discussion}

\begin{figure*}
\centering
 \includegraphics[width=17cm]{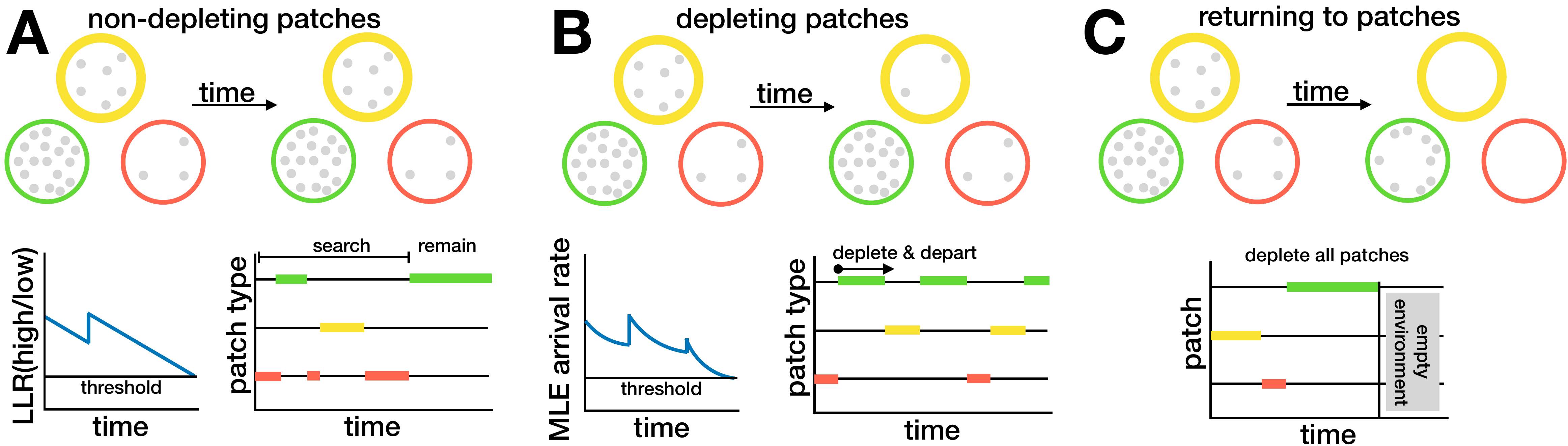} 
\caption{\textbf{Summarized taxonomy of foraging strategies}. See Table~\ref{table:summ} for details. {\bf A.}~In non-depleting environments, an ideal forager searches patches until finding and remaining in a high yielding patch. {\bf B.}~In environments with depleting patches, an ideal forager depletes a patch and departs when reward arrival rate reaches a threshold. {\bf C.}~In fully depleting environments, an ideal forager should fully deplete patches before departing in order to minimize the time to clear the environment.}
\label{fig10:stratsumm}
\end{figure*}

\begin{table*}[t]  
% note:  putting \strut at the beginning and end of each parbox fixes spacing issues https://tex.stackexchange.com/questions/230073/vertical-spacing-between-parboxes
\footnotesize
	\begin{tabular}{p{3cm}p{2mm}p{6cm}p{2mm}p{8cm}} 
	 \hline
		\textit{Environment} & & \textit{Decision strategy/solution} & & \textit{Dependencies/Observations} \\
		\hline \\[-8pt]
		\multicolumn{5}{l}{\parbox{\linewidth}{\strut \raggedright \textbf{Non-depleting patches} \\Objective: Minimize time to arrive and remain in highest yielding patch.  Known:  Food arrival rates of each patch type. \strut}} \\ 
		\hline
		\parbox{\linewidth}{\raggedright $N$-patch types} & &
		\parbox{\linewidth}{\strut \raggedright Depart patch when likelihood of being in highest yielding patch falls below a threshold.\\ $N=2$: See Eqs.~(\ref{binnodep}) \& (\ref{fullcritthet}) and Fig.~\ref{fig2:binnodep_scheme}B;\\ $N=3$: See Eqs.~(\ref{ternnodep}) \& (\ref{terncurvebound}) and Fig.~\ref{fig4:ternenv}B;\\ $N>3$: See Eqs.~(\ref{Nnodepllr}) and Fig.~\ref{fig5:moreno}B. \strut} & &
		\parbox{\linewidth}{\strut \raggedright \textbullet \ Optimal strategy and arrival time depend on fraction of low versus high yield patches and discriminability between 1st/2nd highest yield patches (Figs.~\ref{fig3:binnodep_stats}-\ref{fig5:moreno}).}
		\\ \hline
		\parbox{\linewidth}{\strut \raggedright Continuum of patch types} & 
		& \parbox{\linewidth}{\strut \raggedright Categorize patches as high or low yielding and depart patch if likelihood of a high patch is below threshold (See Eq.~(\ref{rhocont}) and Fig.~\ref{fig5:moreno}D). \strut} & &
		\parbox{\linewidth}{\strut \raggedright \textbullet \ Time to arrive and remain in high yielding patch is non-monotonic in departure threshold, and much longer when high patches are rare. \strut} 
		\\
		\hline \hline \\[-8pt]
		\multicolumn{5}{l}{\parbox{\linewidth}{\strut \raggedright\textbf{Depleting}\\ Objective: Maximize mean food intake rate $R$ over a long time (several patches). Known:  Initial yield rates of each patch type.  }}	
		\\ \hline
		\parbox{\linewidth}{\strut \raggedright 1-patch type} & &
		\parbox{\linewidth}{\strut \raggedright Depart when arrival rate $\lambda (t)$ falls to a threshold value $\lambda_{\theta}$.} & &
		\parbox{\linewidth}{\strut \raggedright \textbullet \ Matches MVT (Eqs.~(\ref{proxRh}) \& (\ref{mvtcp}) and Fig.~\ref{fig6:deplete}B) except when there are very few chunks per patch (Fig.~\ref{fig6:deplete}C) in which case the forager should empty the patch. \strut} 
		\\ \hline
		\parbox{\linewidth}{\strut \raggedright 2-patch types:  patch type known on arrival} & & \parbox{\linewidth}{\strut \raggedright Depart when current patch arrival rate $\lambda_j(t)$ reaches a threshold. \strut} & &
		\parbox{\linewidth}{\strut \raggedright \textbullet \ Represents the `Depletion-dominated' regime, which recovers MVT (See Eqs.~(\ref{bindepknow}) \& (\ref{bindepcp}) and Fig.~\ref{fig6:deplete}D). \strut}  % where the agents stays in the patch long enough to infer type of high/low before leaving}
		\\ \hline
		\parbox{\linewidth}{\strut \raggedright 2-patch types: empty low yield patch} & & \parbox{\linewidth}{\strut \raggedright Wait a time $T_\theta$, then depart patch if no food is found.  If encounter food at $t<T_\theta$, then use threshold on inferred yield rate to make leaving decision (similar to single patch type case) \strut} & & \parbox{\linewidth}{\strut \raggedright \textbullet \  `Uncertainty-dominated' regime deviates from MVT. \\ \textbullet \  Optimal wait time and departure threshold $\lambda_{\theta}$ increase with prevalence of high yielding patch (Fig.~\ref{fig6:deplete}E,F) }
		\\ \hline
		\parbox{\linewidth}{\strut \raggedright 2-patch types: both high and low patches have food} & & \parbox{\linewidth}{\strut \raggedright Decision via threshold on current estimated yield rate $\tilde{\lambda}(t)$ (See Eq.~(\ref{lamtildep})). Choose optimal threshold $\lambda_{\theta}$ that maximizes long term food intake rate.} &  &
		\parbox{\linewidth}{\strut \raggedright \textbullet \ Optimal return differs from known case when there are few food chunks per patch (Fig~\ref{fig7:bindepgen}A), converges to known patch case when there are many chunks per patch (Fig~\ref{fig7:bindepgen}B).\\ \textbullet \ Forager stays in low patches too long, leaves high patches too soon when there are few chunks per patch (uncertainty-dominated regime, Fig~\ref{fig7:bindepgen}C).\strut}
		\\ 
		\hline \hline \\[-8pt]
		\multicolumn{5}{l}{\parbox{\linewidth}{\strut \raggedright \textbf{Returning to patches}\\Objective: Minimize time to harvest all food from the environment. Known:  Yield rates of each patch type. \strut}}		
		\\ \hline
		\parbox{\linewidth}{\strut \raggedright Patch type known on arrival} & & \parbox{\linewidth}{\strut \raggedright Empty each patch completely before moving to the next one. \strut} & & \parbox{\linewidth}{\strut \raggedright \textbullet \  Time to clear environment increase linearly with number of patches in environment (Fig.~\ref{fig8:depenv}F). \strut} 
		\\ \hline
		\parbox{\linewidth}{\strut \raggedright 2-patches:  1 chunk}  & & \parbox{\linewidth}{\strut \raggedright Leave current patch if food not found within waiting time of $T_\theta$ (search problem). \strut} & & \parbox{\linewidth}{\strut \raggedright \textbullet \  Optimal waiting time decreases with discovery rate $\rho$ (Fig~\ref{fig8:depenv}C) in contrast to case without returns (Fig.~\ref{fig6:deplete}E). \strut} 
		\\ \hline
		\parbox{\linewidth}{\strut \raggedright 2-patches: general case} & & \parbox{\linewidth}{\strut \raggedright Consume $m_0^L$ chunks from first patch, then threshold belief to decide patch departure.} & & \parbox{\linewidth}{\strut \raggedright \textbullet Time to clear environment depends weakly on chunk count (Fig.~\ref{fig8:depenv}D) and decreases slightly as chunk encounter rate increases (Fig.~\ref{fig8:depenv}E). \strut}
		\\ 
		 \hline
	\end{tabular}
	\caption{ \textbf{Detailed taxonomy of patch foraging strategies}.  Patch decision strategies depend on the environment, and trends in observables differ for the foraging environments considered in the model.  Environments considered include non-depleting patches, depleting patches, and the case of returning to specific patches to fully deplete an environment (see also Fig \ref{fig10:stratsumm}).  Columns describe important aspects of the optimal decision strategy for each case, along with key model results. 
% JD:  I Removed last two sentences, since  they seem a bit out of context:
%For the different cases, experimental observables include patch residence time, travel time, resource consumption over time, and patch yield rate over time. In addition to these, model-derived observables include the patch type belief threshold $\theta$, time to arrive and stay in a high yielding patch ($T_{\rm arrive}$), patch yield rate threshold $\lambda_{\theta}$, and patch waiting time $T_\theta$.	
	}
	\label{table:summ}
\end{table*}

%%%%%%% SUMMARY %%%%%%%%%%%%%%%%%%%%%%%%%%%%%%%%%%%%%%%%%%%%%%%%%%%%%%%%%%%%%%%%
\subsection{Summary}
Patch leaving decisions are an essential component of foraging. Our model uses principles of probabilistic inference to establish a normative theory of patch leaving decisions, as well as a framework for learning about resource availability in the environment. 

The model yields several key takeaways concerning optimal decision strategies in a variety of patch foraging contexts (see Fig.~\ref{fig10:stratsumm} for an overview and Table \ref{table:summ} for details). In the idealized case in which foraging does not deplete patches, the optimal strategy is for a forager to minimize their time to find and remain in the highest yielding patch in the environment. This is accomplished by triggering patch leaving decisions when the log-likelihood ratio for the probability of high return versus other patches falls below a threshold. When a forager depletes patches, we found that across a broad range of environmental parameters the best strategy for maximizing the environmental reward rate is to depart a patch when the in-patch reward rate matches the average return rate of the environment, i.e. corresponding to the marginal value theorm (MVT). However, if there is high uncertainty about the patch type, the forager will stay longer in low yield patches and shorter in high yield patches than predicted by the MVT. When a forager fully depletes their environment, an optimal strategy minimizes the number of transits between patches in order to deplete the environment as quickly as possible. Lastly, we found an observer that learns the rate of food arrival within patches learns more quickly in depleting than non-depleting environments environments. 

Throughout the study, we compute and vary metrics that have parallels used in systems neuroscience such as reaction times and discriminability.  We solved the model equations via a combination of first passage time methods for stochastic processes along with Monte Carlo sampling of jump processes corresponding to foragers' beliefs.  Since the contexts analyzed in our study (Fig.~\ref{fig10:stratsumm}; Table~\ref{table:summ}) can be realized experimentally, our model provides a framework for the formal quantitative analysis of behavior in patch foraging experiments.

%%%%%%% RELATION TO OTHER WORK %%%%%%%%%%%%%%%%%%%%%%%%%%%%%%%%%%%%%%%%%%%%%%%%%%%%%%%%%%%%%%%%
\subsection{Relation to other work}
% BAYESIAN FORAGING
\noindent
\textbf{Animals often approximate Bayesian foraging.}
%intro: In behavioral ecology, many studies have considered a Bayesian approach to patch leaving, asking how an animal can use available information to make optimal foraging decisions~\cite{biernaskie2009bumblebees, f2006simpler, mcnamara2006bayes, olsson1998survival}.
%However, because these studies each considered a narrow range of environmental conditions, it is not clear how the Bayesian approach relates to other models of foraging decisions, and how such an approach may assist in deriving possible neural implementations. To date there has not been a complete normative theory of how animal make patch leaving decisions under a broad class of environmental conditions.
We have treated patch leaving as a statistical inference problem within a Bayesian framework, assuming the animal has knowledge of the transit time between patches, the impact of food depletion on the food arrival rate, and (in the case without learning) the distribution of patch arrival rates.
Our theoretical treatment of patch leaving decisions builds on previous Bayesian models of foraging~\cite{McNamara_Houston_1980,olsson2006bayesian,pierre2008bayesian,olsson1999gaining,van2010state,rodriguez1997density,ellison2004bayesian,mcnamara2006bayes},
experimental studies that ask if animals behave as Bayesians~\cite{alonso1995patch,green1980bayesian,olsson1999gaining2,j2006animals,killeen1996bayesian,nonacs1998patch,valone1989measuring}, and other experiments that show how an animal's prior knowledge about an environment modulates its foraging behavior~\cite{van2003incompletely, klaassen2006optimal, hughes1995effects}.  %removed:  klaassen2007prior, fauchald1999foraging, oaten1977optimal.  Moved some other refs to below section
For example, bumblebees~\cite{biernaskie2009bumblebees} and inca doves~\cite{valone1991bayesian} adjust their foraging strategies in response to the predictability of the environment, as a Bayesian forager would, 
% Doves exploited patches in a manner consistent with prescient foraging when patch quality was temporally predictable. The same individuals exploited patches in a manner consistent with Bayesian foraging when prescient foraging would not be likely because patch quality was temporally unpredictable. 
but other work shows this is not a universal trend~\cite{marschall_foraging_1989}.
Patch leaving decisions may deviate from Bayes optimality due to animals becoming risk-averse in variable environments~\cite{kacelnik1996risky}. Apart from the behavioral significance of priors in guiding animals' foraging behavior, such cases can help us understand the types of probabilistic neural computations performed in uncertain and dynamic environments.

% LIMITATIONS OF MV 
\noindent
\textbf{Inferential decisions can validate the marginal value theorem.} The marginal value theorem (MVT)~\cite{charnov1976optimal} provides a baseline patch depature rule for comparing optimal foraging to ecological behavior~\cite{stephens2008foraging}.
However, the MVT is limited in its applicability, as it classically describes the case of continuous rewards where the forager has perfect knowledge of resource availability, and does not provide a mechanistic account of how a forager reaches a decision to leave a patch.  To address these limitations, Davidson \& El Hady~\cite{davidson2019foraging} developed a generalized mechanistic model of patch leaving decisions with parameters representing the discreteness of reward arrival, belief noise, and environmental averaging.  In this model, patch leaving decisions are due to a threshold crossing process by an accumulator variable.
Parameters can be adjusted to represent a wide range of foraging strategies not limited to the MVT, and specifically addressing environmental uncertainty. 
This model inspired our current work, which demonstrates that optimal patch leaving decisions involve evidence accumulation. We have identified uncertainty in the current reward rate as a key driver of deviations from the MVT.
Previous studies modeling patch departure under discrete rewards~\cite{Rita_Ranta_1998} have formulated mechanistic descriptions of departure decisions~\cite{waage1979foraging,McNamara_1982,driessen1999patch,taneyhill2010patch}, but none have analyzed such a broad range of task conditions to provide a comparative study of strategies.
% removed: haccou1991information,

% \textbf{Speed-accuracy trade-off}.
% The speed-accuracy trade-off is central to animal decision-making~\cite{chittka2009speed}, and has been considered a hallmark of an accumulation process (but see [Cosyne ref Shadlen lab I think]).
% % Evidence accumulation has been thoroughly studied in trained behavior where animal integrate evidence over prolonged timescales before reaching a decision.  In this case easy decision are fast and the difficult decisions are slower reflecting what is called the speed-accuracy tradeoff.  
% However, stay-or-go decisions like foraging do not have the same speed-accuracy tradeoff [some ref - probably Kacelnik]
% In certain cases, such as non-depleting patches, there is a speed-accuracy tradeoff or in case of choosing between different foraging areas. \newline

% COMPARISON TO MULTI-ARMED BANDIT TASK
\noindent
\textbf{Patch foraging as modified multi-armed bandit}.
Patch leaving behavior involves deciding when to leave the current patch being harvested, noting that available rewards in the next patch may be unknown; in the simplest case departures lead to a random sample of the subsequent patch from the environment. In contrast, the multi-armed-bandit (MAB) task~\cite{srivastava_optimal_2013} assumes an agent has some control over which patch (arm) is chosen next, often does not consider patch depletion, and classically ignores switching costs. However, patch leaving and the MAB task can become similar in certain limits. For instance, patch leaving that involves no depletion, zero switching costs, and memory of past locations is somewhat similar to the classic MAB.

To illustrate the differences, consider an environment with two patch types $H$ and $L$, no depletion of patches, no switching cost, and where the forager knows the patch reward rates values $\lambda_H$ and $\lambda_L$ (treated in Figs.~\ref{fig2:binnodep_scheme} \& \ref{fig3:binnodep_stats}).
Within a patch, the forager uses its experience to decide when to leave the current patch, since they may not know the patch type on arrival. If the forager does not retain memory of the locations of specific patches, then they will randomly come upon the next patch of either type $H$ or $L$ according to the statistics of the environment, and must repeat the process of inferring the reward arrival rate.
Now consider a $K$-arm bandit where some proportion have reward probability $\lambda_H$ and some have $\lambda_L$.
At each step, the gambler chooses which of the $K$ arms to exploit, and continually updates its estimate of which patches are $H$ versus $L$, as the space of arms is typically small enough to learn this well.
Even with the assumptions of no depletion and zero switching costs, as formulated these are still different decision problems: to formulate the MAB problem as a patch foraging problem, the gambler would have a choice at each step between the same arm or a randomly chosen different arm.

Regardless, considerations of switching costs, depletion, and uncertain future rewards are central to patch foraging and stay-or-go decisions, and related work with the MAB has considered some of these cases.
Switching costs were considered by~\cite{guha_multi-armed_2009,asawa_multi-armed_1996,banks_switching_1994}, and including these who generally led to remaining at the same arm for longer, similar to patch foraging.  
The problem of depleting rewards is related to the non-stationary bandit, in the specific case where arm reward rates are choice dependent~\cite{besbes_stochastic_2014,koulouriotis_reinforcement_2008}. 
In a patch with depleting rewards, the expected return changes in a specific and predictable way, and therefore can be treated mathematically, as we showed in Section \ref{deplete}. Limited/incomplete memory of the return of each arm is related to the non-stationary bandit, because non-stationarity of return means estimates of the return of previously visited arms slowly becomes more uncertain over time.  Other MAB task designs consider changing conditions: for example, after an abrupt change in the environment (e.g.~\cite{wei_abruptly-changing_2018}), estimates of the return from each arm can be 'reset'~\cite{hartland_multi-armed_2006}.  A complete reset of the choice policy represents no memory of which arm to choose. This is related to the fact that a forager often has knowledge of only the reward rate of the current patch - finding the next patch requires exploration and possibly an uncertain amount of search costs.   Within these contexts, patch foraging is fairly well described by a non-stationary bandit with (possibly uncertain) switching costs, limited memory or discounting of the choice policy of certain arms, and a specific form of non-stationarity. As such, we see our work providing useful insights about efficient strategies for non-stationary MAB tasks.

% LEARNING
\noindent
\textbf{Bayesian learning could be approximated by reinforcement learning}. At best, animals approximate the optimal learning processes we have described mathematically~\cite{ollason1987learning, pietrewicz1985learning, kamil1982learning}. One such approximation is the relative payoff sum (RPS)-learning rule whereby the probability of choosing a patch is proportional to the relative payoff that the respective patch has yielded so far~\cite{regelmann1986learning}, related to a matching law identified in operant conditioning~\cite{davison2016matching}. A reinforcement learning (RL) framework can also be applied to patch leaving decisions, as discussed by Kolling and Adam~\cite{Kolling_Akam_2017}. These authors showed that while purely model-free RL fails to describe common aspects of patch leaving decisions, model-based predictions of expected rewards may be used to decide when to leave a patch, in conjunction with model-free RL that can be used to evaluate the values of different decision strategies across patches.
% Within this framework, the goal is optimize the average amount of reward received over some period of time (hence the distinction of "average reward RL").  

In our theoretical treatment, we assume the forager uses the correct model of in-patch return rates, consisting of stochastic rewards with an underlying constant rate (non-depleting environment) or decreasing rate (depleting environment). Even if a Bayesian forager did not know a priori whether patches were depleting or not, or what the depletion increment was, a hierarchical Bayesian update could be used to learn this. In our work, the forager uses a threshold $\theta$ on the LLR to make patch leaving decisions in non-depleting environments, or a threshold $\lambda_{\theta}$ on the inferred arrival rate to trigger patch depatures in depleting environments.
Although we obtained explicit solutions for the optimal thresholds $\theta^{\rm opt}$ or $\lambda_\theta^{\rm opt}$ by considering values that lead to maximal return, we note that within the framework of~\cite{Kolling_Akam_2017}, an agent could use model-free RL to learn the optimal values of these parameters. More work needs to be done in order to compare Bayesian mechanisms of learning to RL in the context of patch foraging decisions.

\noindent
\textbf{History and memory effects in patch foraging.} Animals performing trained~\cite{akrami2018posterior, raviv2012recent, ashourian2011bayesian} and naturalistic behavior~\cite{koganezawa2009memory, belisle1997effects, kamil_learning_1990} make history-dependent decisions based on the stimuli they have experienced and the responses they have made.
%It is crucial to delineate between memories and history effects: history effects might appear as the dependence of animal's current behavior on the reward history but memory on the other hand might be the mechanism by which these history effects are mediated.
History-dependence is likely mediated by both short and long term memory processes. Memory for food availability is crucial for efficient foraging, for example in starlings~\cite{bateson1995accuracy,cuthill_starlings_1990}. Movement and search patterns of foraging animals are also governed by environmental estimates from memory~\cite{bracis2015memory}. Our theoretical treatment mostly accounts for memory by assuming the forager retains knowledge of their patch's statistics and the statistics of the environment, as agents update their patch type beliefs by sampling patch food resources.

%%%%%%% APPLICATIONS %%%%%%%%%%%%%%%%%%%%%%%%%%%%%%%%%%%%%%%%%%%%%%%%%%%%%%%%%%%%%%%%
\subsection{Experimental applications: Behavior and neuroscience}

\noindent
\textbf{Experimental realizations of foraging behavior.} There is currently a surge of interest in systems neuroscience to study naturalistic behavior as opposed to traditional experimental designs in which animals are trained over long times to perform particular behavioral tasks (`trained behavior').  One important distinction between the type of foraging behavior we are studying here and trained decision making tasks in perceptual decision making is that the former is continuous, without a specific trial structure, while the latter have a (repeated) trial structure predetermined by experimenters.
The theoretical framework in this work will help experimenters move from trial based behavior to continuous behavior without the loss of quantitative rigor. Studying behavior in this continuous paradigm offers important advantages over trial based paradigms as it allows multi-sensory integration to unfold, increases the fraction of time the animal is engaged in a task, engages the animal's sensory-cognitive-motor loops activated by the environment's natural statistics, and allows neural dynamics to unfold over multiple timescales~\cite{huk2018beyond}. 
%In designing the experiments in a lab to mimic decision making in natural environments, one can control the energy available by providing a controlled amount of resources in the foraging space.  Therefore designing behavioral paradigms that approximate naturalistic foraging incentivizes the subject better and provides more insight into naturalistic decision strategies.

Although recent work has examined foraging tasks within a systems neuroscience framework~\citep{barack2017posterior,hayden2011neuronal, lottem2018activation, vertechi2020inference}, there remain many open questions related to how the brain processes key aspects of foraging decisions.
For example, experiments with the “Self-control preparation,” where an animal chooses between two choices, have significant behavioral differences with those that have a sequential foraging “patch preparation”, even though from an economic standpoint, the setups are equivalent~\cite{stephens2008decision}.
Moreover, many tasks do not allow the animal to move extensively in space, which is a crucial aspect of foraging. For instance, when rats are required to physically move to perform foraging, the observed behavior differs from tasks that “simulate” foraging by presenting sequential choices or that consider visual search~\citep{wikenheiser2013subjective}. 
%Animals that can move during tasks exhibit more complex strategies, and search patterns can be varied by tuning the geometric arrangement of the patches in the foraging space.

\noindent
\textbf{Neural implementation of patch foraging decisions.} As we are aiming to provide a conceptual framework for patch foraging experiments in systems neuroscience, a crucial aim of such experiments is to map brain circuits underlying patch foraging decisions. The behavioral conceptual framework we have treated in this study includes behavioral mechanisms underlying patch leaving decisions in different types of environments. By extension, those behavioral mechanisms should map onto neural mechanisms. As in classic perceptual decision-making tasks, sequential updating models provide key insights into the types of neural  computations one should expect to see in subjects performing those decisions~\cite{gold2007neural}.

Fitting our behavioral model to both the behavioral data and the neural data simultaneously~\cite{batty_behavenet_2019} can unravel a series of intricate neural computations underlying particular behavioral strategies. To date, only a few brain areas have been shown to be involved in different aspects of foraging decisions: ventromedial prefrontal cortex encodes values of well-defined options, and anterior cingulate cortex encodes the average value of the foraging environment and cost of foraging~\cite{kolling2012neural} or signals changes in environmental context~\cite{shenhav2014anterior, shenhav2013expected}. Although experimentalists have traditionally targeted selected stereotypical areas, the increased availability and versatility of large scale recording techniques will enable future experiments to map the global circuits underlying foraging decisions, thus accessing a brain wide representation of intricate behavioral strategies~\cite{kaplan2020brain}. 

\noindent
\textbf{Comparative patch foraging behavior.} Comparative approaches, identifying similarities and differences across species, are a key aspect of the behavioral ecology of foraging~\cite{roubik1981comparative, bonaccorso2007evidence,wajnberg2003comparative}.   Our model enables quantitative, model-based comparisons of decisions strategies between animals, which will improve our understanding of the aspects of evolution (e.g., selection pressure) shaping foraging behavior~\cite{raine2006adaptation, hills2006animal}. Since neural circuits are shaped by these pressures too, this opens novel avenues for comparative systems neuroscience and an evolutionary understanding of neural computations during foraging.

%%%%%%% FUTURE WORK %%%%%%%%%%%%%%%%%%%%%%%%%%%%%%%%%%%%%%%%%%%%%%%%%%%%%%%%%%%%%%%%
\subsection{Model extensions}
Our model forms the foundation for multiple avenues of future theoretical study. One extension, discussed above, concerns using reinforcement learning to tune the patch leaving decision thresholds in the model. While we assumed the agent learns the qualities of different patch types, we fixed the transit time between patches; another extension could focus on learning and decision making in environments with distributed transit times. Such a feature could arise via spatial arrangements of patches, with potentially correlated resource availabilities across adjacent patches~\cite{van2010state,srivastava2015correlated}. 

Effective search is integral to survival in nature~\cite{nathan2008emerging}, and search behavior can give information on individual decision strategies. For example, a recent study found that rats can solve the stochastic travel-salesman problem using a nearest neighbour algorithm~\cite{jackson2019many}. A similar approach could ask how an animal's search and navigation pattern interact with different patch leaving decisions to create an effective foraging strategy.

Our framework has also developed strategies assuming the forager utilizes constant thresholds on either the belief or estimated arrival rate. However, it is well known that in a variety of decision making contexts, optimal strategies can involve time-dependent decision thresholds~\cite{cisek2009decisions,drugowitsch2012cost,drugowitsch2014optimal,malhotra2018time}. Typically, these results arise in the context of multi-trial experiments in which the quality of evidence on each trial varies stochastically and is initially unknown. In the non-depleting environments of our study, the quality of evidence is fixed across patch visits and beliefs are used to trigger departure decisions. As such, models in this context fit the assumptions of classic, constant threshold optimal policies. On the other hand, in uncertain binary depleting environments, we have projected a higher-dimensional description of the patch value to a single scalar estimate of the patch arrival rate. As such, we cannot be sure the parameterized model of patch departure decisions is optimal across all models. Leveraging methods from dynamic programming commonly used to set optimal decision policies would be a fruitful next step in ensuring the optimality of our patch leaving decision strategies.

We defined optimal patch foraging as maximizing resource intake.  However, an alternative formulation could consider an explicit trade‐off between gathering resources (exploitation) and gathering information about the environment (exploration).  Information foraging frames an agent's movement strategy by prioritizing the value of information over resource gathering~\cite{vergassola2007infotaxis}.  In our problem, information foraging would target strategies that reduce the time to learn the environmental distribution of patch qualities (as in Section~\ref{sec:learn}), rather than maximize the reward rate, thus prioritizing the role of exploration and sampling in foraging~\cite{nonacs1998patch, naef2000patch, stephens1982stochasticity, dow1987sampling, kramer1991exploration}.

Although the natural world typically offers a continuous foraging space, considering discrete resource patches facilitates understanding by systematically dissecting fine scale dynamics of local search from global search~\cite{levin2012patch, grunbaum2012logic}. Additionally, animals seems to exploit local patches of food resources as if they had partitioned the world into patches.  Nonetheless, not all resources exist in patches; a further extension of our theoretical framework could consider a continuous world without patches. Assuming some regularity in the distribution of food, it would then be efficient to consider stochastic versions of gradient ascent so the forager could orient itself in the direction of higher food concentrations.

Another important extension will be to consider interactions between agents, either through predator-prey interactions which affect foraging decisions~\cite{greene1986patterns}, social foraging of groups collectively foraging~\cite{clark1984foraging,giraldeau2018social}, or even competitive foraging~\cite{harper1982competitive}.  In our model, the agent only receives direct (non-social) information about resource availability;  in collective foraging, an individual receives both social and non-social information~\cite{templeton1996vicarious,perez-escudero_collective_2011}.  This can significantly affect foraging decisions, for example when an individual must balance a desire to obtain more resources with a desire to maintain group cohesion.

To conclude, our model establishes a formal framework for a natural behavior (patch foraging) that can be studied in the same formal rigor as many trained behavioral tasks.  
Guided by the philosophy that formal tractability of a behavior entails neural mechanistic tractability, future work will build on this framework to generate testable hypotheses on the neural mechanistic underpinnings of foraging behavior.

\begin{acknowledgments}
We thank Jan Drugowitsch for helpful discussions regarding the optimality of decision threshold policies across patch visits. ZPK was supported by a CRCNS grant (R01MH115557) and NSF (DMS-1853630). AEH acknowledges support by NIH grant 1R21MH121889-01.
JDD acknowledges support by the Deutsche Forschungsgemeinschaft (DFG, German Research Foundation) under Germany’s Excellence Strategy - EXC 2117 - 422037984, as well as support from the Heidelberg Academy of Science.

\end{acknowledgments}

%\appendix

\bibliographystyle{unsrt}
\bibliography{normforage}% Produces the bibliography via BibTeX.
\end{document}